\documentstyle[12pt]{article}
\def\SGMPcomment#1{}
\def\SGMPdefifnonempty#1#2{\def\tempcs{#2}%
	\ifx \tempcs\empty \else
		\def#1{#2}%
	\fi
}
\let\mdef=\SGMPdefifnonempty
\newif\ifSGMPkeeplooping

\newtoks\SGMPttlpgtoks	\SGMPttlpgtoks={}
\long\def\SGMPttlpgmisc#1{\relax
	\expandafter\SGMPttlpgtoks\expandafter{\the\SGMPttlpgtoks
				\vskip6pt plus6pt minus4pt#1\par}%
}

\def\SGMPendpreface{%
	\end{titlepage}\global\advance\count0 by\pgn
	\global\advance\count0 by 1
}
\catcode`\@=11
\def\SGMPmaketitle{%
	\gdef\@thanks{\begin{center}\normalsize\the\SGMPttlpgtoks\end{center}}%
	\maketitle
}
\def\SGMPref#1{%
	\@ifundefined{r@x:#1}{\pageref{pg:#1}}{\ref{x:#1}}%
}
\catcode`\@=12
\def\SGMPunident{}
\def\SGMPlab#1{\ifx#1\SGMPunident\else\label{x:#1}\fi}

\def\SGMPbeginList#1#2{%
	\def\SGMPlisttype{#1}%
	\ifx \SGMPlisttype\SGMPlisttext
		\def\SGMPendList{\end{description}}%
		\def\next{\begin{description}}%
		\def\SGMPitem{\item[#2]}%
	\else \ifx \SGMPlisttype\SGMPlistnone
		\def\SGMPendList{\end{description}}%
		\def\next{\begin{description}}%
		\def\SGMPitem{\item[]}%
	\else \ifx \SGMPlisttype\SGMPlistbulleted
		\def\SGMPendlist{\end{itemize}}%
	\else \ifx \SGMPlisttype\SGMPlistsquare
		\def\SGMPendlist{\end{itemize}}%
	\else
		\def\SGMPendList{\end{enumerate}}%
		\def\next{\begin{enumerate}}%
		\def\SGMPitem{\item}%
	\fi \fi \fi \fi
	\next
}
\def\SGMPlisttext{text} \def\SGMPlistnone{none}
\def\SGMPlistbulleted{bulleted} \def\SGMPlistsquare{square}

\makeindex
\def\SGMPindex#1#2{\def\tempcs{#1}%
	\ifx \tempcs\SGMPvalueyes
		#2%
	\fi
	\index{#2}%
}
\def\SGMPvalueyes{yes}
\begingroup \catcode`\@=11
\newread\dfe@ \gdef\dfe#1#2#3{\relax
       \immediate\openin\dfe@=#1 \ifeof\dfe@#3\else#2\fi
       \immediate\closein\dfe@}
\gdef\SGMPstartindex{\relax\ifx\@indexfile\undefined\else
       \closeout\@indexfile \fi\begin{theindex}
       \def\indexentry##1##2{\item##1 ##2}}
\gdef\SGMPfinishindex{\dfe{\jobname.ind}{\def\next{\input
       \jobname.ind}}{\let\next=\relax}\ifx\next\relax \dfe
       {\jobname.idx}{\def\next{\input \jobname.idx}}{\relax}\fi
       \next \end{theindex}}
\endgroup

\def\aalign#1{\leavevmode\vbox{\baselineskip=0pt \lineskiplimit.25ex
  \ialign{##\crcr#1\crcr}}}
\def\SGMPring#1{\aalign{\hidewidth\char"17\hidewidth\cr\noalign{\kern-1.2ex}#1}}

\let\SGMPnewline=\\

\newtoks\TexMacPairEndtextoks

\def\SGMPgobble#1{}

\def\SGMPlim#1{\def\tempcs{#1}%
	\ifx \tempcs\empty
		\let\SGMPdolim=\displaylimits
	\else \if #1c
		\let\SGMPdolim=\limits
	\else \if #1r
		\let\SGMPdolim=\nolimits
	\else
		\let\SGMPdolim=\relax
	\fi \fi \fi
}

\def\Rad#1{%
	\begingroup
	\def\RadTempCs{{#1}}\let\RdxTempCs=\empty
}

\def\DoRad{%
	\relax
	\ifx \RdxTempCs\empty
		\sqrt\RadTempCs
	\else
		\root \RdxTempCs \of \RadTempCs
	\fi
	\endgroup
}

\def\LeftPost#1{\csname LP#1\endcsname}
\def\RightPost#1{\csname RP#1\endcsname}

\def\getchar #1#2\endgetchar{\def\gotchar{#1}\def\ungotchars{#2}}

\def\SGMPmathgrk#1{%
    \def\ungotchars{#1}%
    \SGMPkeeploopingtrue
    \loop
	\expandafter\getchar\ungotchars\endgetchar
	\ifx \gotchar\empty \def\gotchar{0}\fi
	\count255=\expandafter`\gotchar\relax
	\advance\count255 by -49
	\ifcase \count255
		\nabla
	\or	\varpi
	\or	\varepsilon
	\or	\varphi
	\or		
	\or	\partial
	\or\or
	\or	\varrho
	\or\or\or\or\or\or\or
	\or	A%
	\or	B%
	\or	X%
	\or	\Delta
	\or	E%
	\or	\Phi
	\or	\Gamma
	\or	H%
	\or	I%
	\or		
	\or	K%
	\or	\Lambda
	\or	M%
	\or	N%
	\or	O%
	\or	\Pi
	\or	\Theta
	\or	P
	\or	\Sigma
	\or	T%
	\or	\Upsilon
	\or
	\or	\Omega
	\or	\Xi
	\or	\Psi
	\or	Z%
	\or\or\or\or\or\or
	\or	\alpha
	\or	\beta
	\or	\chi
	\or	\delta
	\or	\epsilon
	\or	\phi
	\or	\gamma
	\or	\eta
	\or	\iota
	\or	\vartheta
	\or	\kappa
	\or	\lambda
	\or	\mu
	\or	\nu
	\or	o%
	\or	\pi
	\or	\theta
	\or	\rho
	\or	\sigma
	\or	\tau
	\or	\upsilon
	\or	\varsigma
	\or	\omega
	\or	\xi
	\or	\psi
	\or	\zeta
	\else
	\fi
	\relax
	\ifx \ungotchars\empty \SGMPkeeploopingfalse \fi
	\ifSGMPkeeplooping
    \repeat
}

\catcode`\@=11
\def\eqalign#1{\null\,\vcenter{\openup\jot\m@th
  \ialign{\strut\hfil$\displaystyle{##}$&$\displaystyle{{}##}$\hfil
      \crcr#1\crcr}}\,}
\catcode`\@=12

\def\MthAcnt#1#2{#2{#1}}

\def\SGMPgraphic#1#2#3#4#5#6#7{{%
	\def\type{#3}
	\def\imresdefault{#4}
	\def\imresvalue{#5}
	\def\picresdefault{#6}
	\def\picresvalue{#7}

	\def\yes{yes}
	\def\drawing{drawing}
	\def\image{image}
	\def\height{4in}

	\ifx\type\drawing	
	    \vbox to\height{%
			\special{pub: pubdraw #2 #10}
			\vfil}
	\else\ifx\type\image	
	    \ifx\imresdefault\yes
		\vbox to\height{%
			\vfil
			\special{pub: sunbitmap #2 #10 0}}
	    \else
		\vbox to\height{%
			\vfil
			\special{pub: sunbitmap #2 #10 \imresvalue}}
	    \fi
	\else			
	    \ifx\picresdefault\yes
		\vbox to\height{%
			\vfil
			\special{pub: sunbitmap #2 #10 0}}
	    \else
		\vbox to\height{%
			\vfil
			\special{pub: sunbitmap #2 #10 \picresvalue}}
	    \fi
	\fi\fi
}}

\ifx\TMPcountA\undefined
  \catcode`\!=10
\else
  \catcode`\!=14
\fi
!\newcount\TMPcountA
!\newcount\TMPcountB
!\newdimen\TMPdimenA
!\newdimen\TMPdimenB
\catcode`\!=12

\def\postscript#1#2#3#4#5#6{%
  \TMPdimenA=#5\relax
  \TMPdimenB=#6\relax
  \TMPcountA=\TMPdimenA
  \TMPcountB=\TMPdimenB
  \hbox to #1{%
    \vbox to #2{
      \vss
      \special{ps: plotfile #3 asis}
      \special{ps::[asis,end]
         ChartCheckPoint restore
         0 SPE
      }
    }%
    \hss
  }%
}

\def\SGMPTabcnvtlist#1#2{%
	\def\tempcsA{#1}%
	\def#2{}%
	\ifx \tempcsA\empty \else
	    \SGMPkeeploopingtrue
	    \loop
		\expandafter\SGMPparsetablelist\tempcsA:::\endSGMPparsetablelist#2
		\ifx \tempcsA\empty
			\SGMPkeeploopingfalse
		\fi
		\ifSGMPkeeplooping
	    \repeat
	\fi
}
\def\SGMPparsetablelist #1:#2::#3\endSGMPparsetablelist#4{%
	\def\tempcsA{#2}%
	\expandafter\def\expandafter#4\expandafter{#4\\#1}%
}

\def\SGMPTabColW#1{\SGMPTabcnvtlist{#1}\TabColW}

\def\SGMPTableWd#1{\def\tempcs{#1}%
	\ifx \tempcs\SGMPabs
		\def\TableWd{A}%
	\else
		\def\TableWd{R}%
		\def\TableWdRPct{#1}%
	\fi
}

\def\SGMPabs{abs}

\def\SGMPbeginTable#1#2#3#4#5#6#7#8#9{%
	\edef\TabRuleVO{\TabRuleVI}\edef\TabRuleHO{\TabRuleHI}%
	\SingleRuleWidthInPixels=6
	\Table[\SGMPTableWd{#9}\mdef\TableJust{#6}\SGMPTabColW{#5}%
		\mdef\TabJustVO{#8}\mdef\TabJustVH{#8}%
		\mdef\TabRuleHI{#7}\mdef\TabRuleHO{#7}\mdef\TabRuleHH{#7}%
		\mdef\TabRuleVI{#3}\mdef\TabRuleVO{#3}%
		\mdef\TabJustHO{#1}\mdef\TabJustHH{#1}%
		\SGMPTabJustHS{#4}%
		\SGMPTabRuleVS{#2}]%
	\let\SGMPnewline=\newline
}

\def\newline{\relax
	\ifvmode
		\vskip\baselineskip
	\else
		\unskip\vadjust{}\nobreak\hfil\break\vadjust{}\ignorespaces
	\fi
}

\def\SGMPTabJustHS#1{\SGMPTabcnvtlist{#1}\TabJustHS}

\def\SGMPTabRuleVS#1{\SGMPTabcnvtlist{#1}\TabRuleVS

\expandafter\SGMPrminitialdblsh\TabRuleVS\\\\\\\endSGMPrminitialdblsh\TabRuleVS
}

\def\SGMPrminitialdblsh\\#1\\\\#2\endSGMPrminitialdblsh#3{\def#3{#1}}

\begin{document}

\vskip 3cm

\centerline{\bf QUANTUM EFFECTS IN COSMOLOGY}
\vskip 1cm
\centerline{\bf L.P. GRISHCHUK}
\centerline{Sternberg Astronomical Institute, Moscow University}
\centerline{119899 Moscow V-234, Russia}
\centerline{and}
\centerline{Physics Department, Washington University}
\centerline{St. Louis, 63130, U.S.A.}

\newpage

\par \underline{{\protect\normalsize {\bf Contents}}}\par

\SGMPbeginList{numeral}{}

\SGMPitem\def\XRefId{}\SGMPlab\XRefId Introduction.
The Present State of the Universe.

\SGMPitem\def\XRefId{}\SGMPlab\XRefId What Can We Expect
 From a Complete Cosmological Theory?

\SGMPitem\def\XRefId{}\SGMPlab\XRefId An Overview
of Quantum Effects in Cosmology.

\SGMPitem\def\XRefId{}\SGMPlab\XRefId Parametric (Superadiabatic)
Amplification of Classical Waves.

\SGMPitem\def\XRefId{}\SGMPlab\XRefId Graviton Creation
in the Inflationary Universe.

\SGMPitem\def\XRefId{}\SGMPlab\XRefId Quantum States
of a Harmonic Oscillator.

\SGMPitem\def\XRefId{}\SGMPlab\XRefId Squeezed Quantum States
of Relic Gravitons and Primordial Density Perturbations.

\SGMPitem\def\XRefId{}\SGMPlab\XRefId Quantum Cosmology,
Minisuperspace Models and Inflation.

\SGMPitem\def\XRefId{}\SGMPlab\XRefId From the Space
of Classical Solutions to the Space of Wave Functions.

\SGMPitem\def\XRefId{}\SGMPlab\XRefId On the Probability
of Quantum Tunneling From {``}Nothing{''}.

\SGMPitem\def\XRefId{}\SGMPlab\XRefId Duration of Inflation
and Possible Remnants of the Preinflationary Universe.

\SGMPitem\def\XRefId{}\SGMPlab\XRefId Relic Gravitons
and the Birth of the Universe.

\SGMPendList

\noindent
Acknowledgements\SGMPnewline
References\SGMPnewline
Figures and Figure Captions\par

\newpage

\par \underline{{\protect\normalsize {\bf 1.
Introduction {--} The Present State of the
Universe}}}\par

\par The contemporary classical cosmology successfully describes the
main features and evolution of the universe, but uses for this some
specific initial conditions. In the framework of classical cosmology
these conditions do not have their own reasonable explanation. They are
just selected in such a way that the theoretical predictions be
compatible with the actual observations. A more deeper understanding of
why the  universe has these and not the other properties can be provided
by quantum cosmology. The most important unsolved issue is the nature of
the cosmological singularity whose existence follows from the classical
general relativity. The phenomenon of singularity is probably the most
compelling reason for  replacing classical cosmology with quantum
cosmology.\par

\par Let us recall some properties of the present world which seem to
have their origin in the very early  universe (see standard textbooks
[1]).\par

\par The distribution of galaxies in space as well as the distribution
of their red shifts indicate that at the largest accessible scales the
universe is more or less homogeneous and isotropic. The most convincing
manifestation of the large scale homogeneity and isotropy of the
universe is the very low level of the angular variations of the
temperature of the microwave background radiation:
\(
\SGMPmathgrk{D}T/T{\ifmmode\approx\else$\approx$\fi}5\cdot10^{-6}\)
 [2]. All the observational data point out to the
conclusion that the overall structure and dynamics of the observable
part of the universe can be described, up to small perturbations, by the
Friedmann (or Friedmann-Robertson-Walker, FRW) metrics:
\def\XRefId{}
\begin{equation}\SGMPlab\XRefId\vcenter{\halign{\strut\hfil#\hfil&#\hfil\cr
$\displaystyle{ds^{2}=c^{2}dt^{2}-a^{2}{\left\LeftPost{par}
t\right\RightPost{par}}dl^{2}.}$\cr
}}\nonumber\end{equation}

It is known that
\(
dl^{2}\)
, the spatial part of the metrics
\(
{\left\LeftPost{par}1\right\RightPost{par}}\)
, can correspond to the closed
\(
{\left\LeftPost{par}k=+1\right\RightPost{par}},\)
 open
\(
{\left\LeftPost{par}k\hskip 0.167em =\hskip 0.167em -\hskip 0.167em
1\right\RightPost{par}}\)
 or flat
\(
{\left\LeftPost{par}k\hskip 0.167em =\hskip 0.167em 0\right\RightPost{par}}
\)
 3-dimensional spaces. The actual sign of the spatial
curvature depends on the ratio
\(
\SGMPmathgrk{W}=\hskip 0.167em \SGMPmathgrk{r}_{m}/\SGMPmathgrk{r}_{c}
\)
 of the mean total matter density
\(
\SGMPmathgrk{r}_{m}\)
 to the critical density
\(
\SGMPmathgrk{r}_{c}={{3}\over{8\SGMPmathgrk{p}G}}H^{%
2}\)
, where
\(
H\)
 is the Hubble parameter
\(
H{\left\LeftPost{par}t\right\RightPost{par}},\hskip 0.212em
H{\left\LeftPost{par}
t\right\RightPost{par}}\equiv \MthAcnt {a}{\dot }\hskip 0.167em
\hskip 0.167em /a.\)
 The majority of the available astronomical data
favour the value $\Omega${\ifmmode<\else$<$\fi}1 which implies that
\(
k=\hskip 0.167em -\hskip 0.167em 1.\)
 However the observations can not presently exclude
neither
\(
k=\hskip 0.167em 0\)
 nor
\(
k=\hskip 0.167em +1\)
. In any case, the current value of the parameter
$\Omega$ is close to one.\par

\par Although the overall structure of the  universe is homogeneous and
isotropic, it is obviously inhomogeneous and anisotropic at scales
characteristic for galaxies and their clusters. It is believed that
these inhomogeneities were formed as a result of growth of small initial
(primordial) perturbations. In order to produce the observed
inhomogeneities the initial perturbations must have had the specific
amplitude and specific spectrum. There are some theoretical and
observational arguments in support of the so-called {``}flat{''}
Harrison-Zeldovich spectrum [3] of the initial fluctuations. In order to
be compatible with the observations this picture may also require a
significant amount of {``}dark{''} matter.\par

\par The dynamical characteristics of the averaged distribution of the
matter, the growth and formation of the small scale inhomogeneities, the
abundances of various chemical elements, as well as other features of
the actual universe, are all successfully brought together by the
{``}standard{''} classical cosmological theory. The trouble is,
however, that the {``}standard{''} theory postulates certain
properties of the  universe rather than derives them from more
fundamental principles. For instance, the observational fact of the
angular uniformity of the temperature of the microwave background
radiation over the sky does not have any rational explanation except of
being a consequence of the postulated, everlasting homogeneity and
isotropy (plus small perturbations). A more natural explanation to a set
of observational facts can be provided by the inflationary hypothesis
[4]. Of course, this hypothesis has its own limits of applicability and
conditions of realization, and after all it may prove to be wrong, but
the phenomenon of inflation seems to be quite general and stable. This
is why it is worthwhile to investigate its consequences and compare with
observations.\par

\par  According to the inflationary hypothesis the spatial volume of the
universe confined to the current Hubble distance
\(
l_{H\hskip 0.167em }=\hskip 0.167em c/H\hskip 0.167em
{\ifmmode\approx\else$\approx$\fi}2\cdot10^{28}\)
 cm or, possibly, even much larger volume, has
developed from a small region which was causally connected in the very
distant past. If the inflationary stage in the evolution of the very
early universe did really take place,  the large scale homogeneity and
isotropy, as well as the closeness of $\Omega$ to one, can be explained as
the consequences of the inflationary expansion.\par

\par The simplest model for the inflationary stage of expansion is
provided by the De-Sitter solution. Originally it was derived as a
solution to the vacuum Einstein equations with a constant cosmological
$\Lambda$-term. However, it can also be treated as a solution to the
Einstein equations with matter satisfying the effective equation of
state p = -$\epsilon$. The De-Sitter solution describes a space-time with a
constant 4-curvature. This space-time is as symmetric as Minkowski
space-time, in the sense that it also admits the 10-parameter group of
motions. The line element of the De-Sitter space-time has the form (see,
for example, [5]):
\def\XRefId{}
\begin{equation}\SGMPlab\XRefId\vcenter{\halign{\strut\hfil#\hfil&#\hfil\cr
$\displaystyle{ds^{2}\hskip 0.167em =\hskip 0.167em c^{2}
dt^{2}-a^{2}{\left\LeftPost{par}t\right\RightPost{par}}
{\left\LeftPost{sqb}dr^{2}+sin^{2}r{\left\LeftPost{par}d\SGMPmathgrk{q}^{%
2}+sin^{2}\SGMPmathgrk{q}d\SGMPmathgrk{y}^{2}\right\RightPost{par}}
\right\RightPost{sqb}},}$\cr
}}\end{equation}

\noindent
where
\(
a{\left\LeftPost{par}t\right\RightPost{par}}\hskip 0.167em =\hskip 0.167em
r_{0}ch{{ct}\over{r_{0}}}\),
and
\(
r_{0}\hskip 0.167em =\hskip 0.167em {\rm const}\).
It is known that in the De-Sitter space-time one can
also introduce the frames of reference with flat or open (hyperbolic)
space sections, though these coordinate systems do not cover the whole
of the De-Sitter space-time. The often used is the flat 3-space
representation:
\def\XRefId{}
\begin{equation}\SGMPlab\XRefId\vcenter{\halign{\strut\hfil#\hfil&#\hfil\cr
$\displaystyle{ds^{2}\hskip 0.167em =\hskip 0.167em c^{2}
dt^{2}\hskip 0.167em -\hskip 0.167em a^{2}{\left\LeftPost{par}
t\right\RightPost{par}}\hskip 0.167em {\left\LeftPost{par}dx^{2}
+dy^{2}+dz^{2}\right\RightPost{par}},}$\cr
}}\end{equation}

\noindent
where
\(
a{\left\LeftPost{par}t\right\RightPost{par}}=e^{H_{0}t},\)
and
\( H_{0}={\rm const}\).  The constant
\( H_{0}\) is the Hubble parameter at the De-Sitter stage of
expansion. The scale factor
\( a{\left\LeftPost{par}t\right\RightPost{par}}\)
of the line element (2) approaches the behaviour
\(
a{\left\LeftPost{par}t\right\RightPost{par}}\hskip 0.167em =\hskip 0.167em
e^{H_{0}t}\)
very quickly during several characteristic time intervals
\(
t\hskip 0.167em =\hskip 0.167em {{1}\over{c}}
\hskip 0.167em r_{0}\hskip 0.167em .\)
\par

\par In order to see the advantages of the inflationary hypothesis let
us assume that intially, during the Planck era, the distance between two
idealized physical objects was of order of a few Planckian scales,
\(
l_{pl}\hskip 0.167em =\hskip 0.167em 10^{-33}
\)
 cm. It can be shown that the present day distance
between these objects can be as large as the present day Hubble distance
\(
l_{H}\hskip 0.167em {\ifmmode\approx\else$\approx$\fi}2\cdot10^{28}\) cm,
if the duration of inflationary stage
\(
\SGMPmathgrk{D}t\)
 was sufficiently long,
\( H_0 \Delta t > 65 \)
In this way a small causally connected region could
have expanded to the size of the presently observed  universe, and the
uniformity of
\( T\) over the sky could have been established by causal
physical processes in the very early universe. At the same time, a
sufficiently long inflation makes the present value of the
\(
\SGMPmathgrk{W}\)
parameter very close to
\(
\SGMPmathgrk{W}=1\). \SGMPnewline
\par

\par \underline{{ {\bf 2.
What Can We Expect From a Complete Cosmological}}}\newline
\underline{{{\bf
Theory?}}}\par

\par The hypothesis of the inflationary stage helps us to make some of
the cosmological data more {``}natural{''}. However, the origin of
the inflationary stage still needs to be explained. The question still
remains, what  kind of evolution did the universe experience before the
inflationary stage and how did the universe itself originate? A
frequently made assumption is that prior to the De-Sitter stage there
was a preceding radiation-dominated era. This assumption just postpones
the answer to the above mentioned questions and returns us to the
problem of cosmological singularity and quantum gravity. As a more
fundamental solution to the problem, it was suggested [6] that the
inflationary era was preceded by an essentially quantum-gravitational
phenomenon called a spontaneous birth of the  universe. A theory capable
of describing the classical stages of evolution of the  universe, as
well as its quantum-gravitational origin, can be named a complete
cosmological theory. Let us speculate on the main expected features of
such a theory.\par

\par The desired evolution of the scale factor
\(
a{\left\LeftPost{par}t\right\RightPost{par}}\)
 is shown in Fig. 1. According to this scenario, the
moment of appearance of the classical  universe corresponded to
\(
t=0.\)
 After that moment of time the inflationary evolution
has started and has been governed by Eq. (2). It is natural to expect
that all the characteristic parameters of the newly born  universe were
of order of the Planckian scales, i.e. the classical space-time came
into being near the limit of applicability of classical general
relativity. The inflationary expansion may be able to pickup such a
micro-universe and to increase its size up, at least, to the present day
Hubble radius. The wiggly line joining the points
\(
a\hskip 0.167em =\hskip 0.167em 0\)
and
\(
a=l_{pl}\)
at Fig. 1 was meant to describe an essentially
quantum-gravitational process similar to the quantum tunnelling or
quantum decay which could have resulted in the nucleation of the
universe in the state of classical De-Sitter expansion. It is reasonable
to suppose that at the beginning of classical evolution the deviations
from the highly symmetric De-Sitter solution were negligibly small.
Moreover, it seems to be sufficient to take these deviations with the
minimally possible amplitude, i.e. at the level of quantum zero-point
fluctuations. During the inflationary period these fluctuations could
have been amplified and produce the density perturbations and
gravitational waves. The density perturbations are needed to form the
observed inhomogeneities in the  universe. Gravitational waves seem to
be the only source of impartial information about the inflationary epoch
and the quantum birth of the universe.\par

\par These matters have been subjects of study in many research papers
by various authors. We will discuss them in more detail below. We will
see that some of the notions introduced above have acquired more precise
formulation and some of the problems have been partially
solved.\SGMPnewline
\par

\par \underline{{\protect\normalsize {\bf 3.
An Overview of Quantum Effects in
Cosmology}}}\par

\par From this brief exposition of a complete cosmological theory it is
clear that quantum effects and quantum concepts should play a decisive
role in different contexts and at different levels of approximation. It
is useful to give a short classification of the areas of further
discussion where the quantum notions will be dealt with. It is worth
emphasizing that we will often use below the common and powerful
technique which is the splitting up of a given problem into the
{``}background{''} and {``}perturbational{''} parts.\par

\par We will start from a description of classical perturbations on a
classical background space-time. The physical meaning of such effects as
parametric amplification of cosmological perturbations, and first of
all, amplification of gravitational waves, can be clearly seen already
at this level of approximation. The next level of approximation treats
the perturbations as quantized fields interacting with the variable
gravitational field of the nonstationary universe, or, in geometrical
language, with the classical background geometry. A particular, but not
obligatory, example of the variable gravitational field is provided by
the inflationary expansion. At this level of approximation, we will see
how the initial vacuum state of the quantized fields evolves into a pure
multiparticle state with very specific quantum properties. It will be
shown that the final quantum state belongs to the class of the so-called
squeezed quantum states. Squeezing is a very distinct feature
potentially allowing to prove or disprove the quantum origin of the
primordial cosmological perturbations.\par

\par At a still deeper level, the background geometry and matter fields
are also treated quantum-\-mechanically {---} this is the realm of
quantum cosmology. The main object of interest in quantum cosmology is
the wave function of the universe which, in general, describes all
degrees of freedom at the equal footing. This level of discussion is
appropriate for tackling such issues as the beginning and the end of
classical evolution as well as quantum birth of the universe. However,
there is no one unique wave function of the universe, there are many of
them. All possible wave functions constitute the whole space of the wave
functions. Presently, we do not know any guiding principle allowing to
prefer one cosmological wave function over others. This is why we are
facing a painful job of analyzing all of them trying to introduce a
probability measure in the space of all wave functions.\par

\par Going still further, one can introduce a notion of a Wave Function
given in the space of all possible wave functions. In other words, a
wave function of the universe becomes an operator acting on the Wave
Function describing the many universes system. This is a subject of the
now popular so-called third-quantized theory. It is aimed at describing
the multiple production and annihilation of the baby-universes. This
fascinating subject is still at the beginning of its development and is
beyond the scope of the present papers. The reader is referred to the
recent review and technical papers on the subject [7]. In some sense the
different theories listed from above to the bottom are various
approximation to the theories listed in the opposite direction.\par

\par Let us start from the classical theory of small perturbations
superimposed on a given background space-time.\SGMPnewline
\par

\par \underline{{\protect\normalsize {\bf 4.
Parametric (Superadiabatic) Amplification of Classical\newline
Waves}}}\par

\par Let us consider classical weak gravitational waves. The main
purpose is to study the parametric (superadiabatic) amplification of
gravitational waves [8, 9]. The same mechanism is applicable to other
fluctuations if they are governed by similar equations.\par

\par We assume that the space-time metric g${}_{\mu\nu}$ can be
presented in the form
\(
g_{\SGMPmathgrk{m}\SGMPmathgrk{n}}\hskip 0.167em
 {\ifmmode\approx\else$\approx$\fi}\hskip 0.167em
g^{{\left\LeftPost{par}0\right\RightPost{par}}}_{\SGMPmathgrk{m}
\SGMPmathgrk{n}}\hskip 0.167em +\hskip 0.167em h_{\SGMPmathgrk{m}
\SGMPmathgrk{n}}\), where
\(
g^{{\left\LeftPost{par}0\right\RightPost{par}}}_{\SGMPmathgrk{m}
\SGMPmathgrk{n}}\)
 is the background metric:
\def\XRefId{}
\begin{equation}\SGMPlab\XRefId\vcenter{\halign{\strut\hfil#\hfil&#\hfil\cr
$\displaystyle{ds^{2}=a^{2}{\left\LeftPost{par}
\SGMPmathgrk{h}\right\RightPost{par}}
{\left\LeftPost{par}d\SGMPmathgrk{h}^{2}-dx^{2}-dy^{2}
-dz^{2}\right\RightPost{par}}}$\cr
}}\end{equation}

\noindent
and
\( h_{\SGMPmathgrk{m}\SGMPmathgrk{n}}\)
are gravitational wave perturbations. The functions
\(
h_{\SGMPmathgrk{m}\SGMPmathgrk{n}}{\left\LeftPost{par}{\bf x},\hskip 0.167em
\SGMPmathgrk{h}\right\RightPost{par}}\)
can be simplified and some of them put to zero,
\(
h_{0\SGMPmathgrk{m}}=0\),
by using the available gauge freedom. The remaining
components can be decomposed into the mode functions, so that for a
given mode one has
\def\XRefId{}
\begin{equation}\SGMPlab\XRefId\vcenter{\halign{\strut\hfil#\hfil&#\hfil\cr
$\displaystyle{h^{k}_{i}={{1}\over{a}}\SGMPmathgrk{m}
{\left\LeftPost{par}\SGMPmathgrk{h}\right\RightPost{par}}\hskip 0.167em G^{%
k}_{i}{\left\LeftPost{par}x,y,z\right\RightPost{par}},}$\cr
}}\end{equation}
where
\(
h^{k}_{i}=h_{ij}g^{{\left\LeftPost{par}0
\right\RightPost{par}}jk}\).\par

\par In the case under discussion the eigenfunctions
\( G^{k}_{i} \)
can be taken in the simplest form:
\(
G^{k}_{i}=\exp {\left\LeftPost{sqb}\hskip 0.167em
{\ifmmode\pm\else$\pm$\fi}i{\left\LeftPost{par}n_{1}x+n_{2}y+n_{3}
z\right\RightPost{par}}\right\RightPost{sqb}}\hskip 0.167em ,n^{%
2}_{1}+n^{2}_{2}+n^{3}_{%
3}=n^{2}.\)
The main equation to be solved is
\def\XRefId{}
\begin{equation}\SGMPlab\XRefId\vcenter{\halign{\strut\hfil#\hfil&#\hfil\cr
$\displaystyle{\SGMPmathgrk{m}''+\SGMPmathgrk{m}{\left%
\LeftPost{par}n^{%
2}-U{\left\LeftPost{par}\SGMPmathgrk{h}\right\RightPost{par}}\right\RightPost{par}}
=0,}$\cr
}}\end{equation}
where
\(
{\ifmmode{}^\prime\else${}^\prime$\fi}\hskip 0.167em =d/d\SGMPmathgrk{h}\),
\(
U{\left\LeftPost{par}\SGMPmathgrk{h}\right\RightPost{par}}=a^{''}/a,
\)
\(
n\)
is the wave-number, and the wavelength is
\(
\SGMPmathgrk{l}=2\SGMPmathgrk{p}a/n. \) \par

\par Equation (6) describes an oscillator with the varying frequency,
that is we are dealing with a parametrically excited oscillator. It is
also worth noting that this equation is mathematically similar to the
spatial part of the Schr\"odinger equation, with the prime playing
the role of a spatial derivative, for a particle with energy
\( n^{2}\) and the potential
\(
U{\left\LeftPost{par}\SGMPmathgrk{h}\right\RightPost{par}}\).
A typical potential barrier
\( U{\left\LeftPost{par}\SGMPmathgrk{h}\right\RightPost{par}}\)
is shown in Fig. 2.\par

\par In the intervals of
\(
\SGMPmathgrk{h}\)-time such that
\(
n^{2}{\ifmmode>\else$>$\fi}{\ifmmode>\else$>$\fi}
{\ifmmode |\else$ |$\fi}U{\left\LeftPost{par}
\SGMPmathgrk{h}\right\RightPost{par}}
{\ifmmode |\else$ |$\fi}\)
the solutions to Eq. (6) have the form
\(
\SGMPmathgrk{m}=e^{{\ifmmode\pm\else$\pm$\fi}in\SGMPmathgrk{h}}\),
so that one has usual high-frequency waves with the
adiabatically changing amplitude: ~
\(
h={{1}\over{a}}\sin {\left\LeftPost{par}
n\SGMPmathgrk{h}+\SGMPmathgrk{f}\right\RightPost{par}}\).
In the expanding world the amplitude decreases. The
amplitudes of the waves with such
\( n\) that
\(
n^{2}{\ifmmode>\else$>$\fi}{\ifmmode>\else$>$\fi}
{\ifmmode |\else$ |$\fi}U{\left\LeftPost{par}
\SGMPmathgrk{h}\right\RightPost{par}}
{\ifmmode |\else$ |$\fi}\)
for all values of
\(
\SGMPmathgrk{h}\)
decrease adiabatically for all
\(
\SGMPmathgrk{h}\).
These waves are shown symbolically by the wavy line
with the decreasing amplitude above the potential barrier
\(
U{\left\LeftPost{par}\SGMPmathgrk{h}\right\RightPost{par}}\)
in Fig. 2.\par

\par If for a given
\( n\)
there is an interval of time when
\(
n^{2}{\ifmmode<\else$<$\fi}{\ifmmode<\else$<$\fi}
{\ifmmode |\else$
|$\fi}U{\left\LeftPost{par}\SGMPmathgrk{h}\right\RightPost{par}}
{\ifmmode |\else$ |$\fi}\),
the solutions to the second-order differential
equation (6) are no longer oscillatory. In case
\(
U{\left\LeftPost{par}\SGMPmathgrk{h}\right\RightPost{par}}=a^{''}/a
\)
they are
\(
\SGMPmathgrk{m}_{1}=a\)
and
\(
\SGMPmathgrk{m}_{2}=a\int a^{-2}d\SGMPmathgrk{h}
\).
The waves satisfying
\(
n^{2}{\ifmmode<\else$<$\fi}{\ifmmode<\else$<$\fi}
{\ifmmode |\else$
|$\fi}U{\left\LeftPost{par}\SGMPmathgrk{h}\right\RightPost{par}}
{\ifmmode |\else$ |$\fi}\)
for some
\(
\SGMPmathgrk{h}\)
encounter the potential barrier and are governed by
the solutions $\mu$${}_{1}$ and $\mu$${}_{2}$ in the
under-barrier region. The amplitude
\(
\SGMPmathgrk{m}_{f}\)
of the function
\(
\SGMPmathgrk{m}{\left\LeftPost{par}\SGMPmathgrk{h}\right\RightPost{par}}\)
right after exit of the wave from under the barrier
depends on the initial phase of the wave $\phi$. The exiting amplitude
\(
\SGMPmathgrk{m}_{f}\)
can be larger or smaller than the entering amplitude
\(
\SGMPmathgrk{m}_{i}\)
defined right before the wave encountered the
barrier. However, averaging of
\(
{\left\LeftPost{par}\SGMPmathgrk{m}_{f}\right\RightPost{par}}^{2
}\)
over the initial phase $\phi$ (i.e., integrating from
0 to
\(
2\SGMPmathgrk{p}\) ) leads always to the dominant contribution from the
solution $\mu$${}_{1}$. This means that the adiabatic factor
\(
1\hskip 0.167em /\hskip 0.167em a\)
is cancelled out by
\(
\SGMPmathgrk{m}_{1}=a\)
and the amplitude
\( h\) (with the factor
\( 1/a\)
taken into account) of a {``}typical{''} wave
can be regarded as remaining constant in the region occupied by the
barrier. It stays constant instead of diminishing adiabatically as the
waves above the potential barrier do. Thus, the exiting amplitude
\( h_{f}\) of a {``}typical{''} wave is equal to the
entering amplitude
\( h_{i}\)
and is larger than it would have been, if the wave
behaved adiabatically (see Fig. 2).\par

\par The amplification coefficient
\(
R{\left\LeftPost{par}n\right\RightPost{par}}\)
for a given
\( n\) is just the ratio
\(
a{\left\LeftPost{par}\SGMPmathgrk{h}_{f}\right\RightPost{par}}\hskip 0.167em
/\hskip 0.167em a{\left\LeftPost{par}\SGMPmathgrk{h}_{i}\right\RightPost{par}}
\)
where
\(
a{\left\LeftPost{par}\SGMPmathgrk{h}_{i}\right\RightPost{par}}\)
is the value of the scale factor at the last
oscillation of the wave before entering the under-barrier region, and
\(
a{\left\LeftPost{par}\SGMPmathgrk{h}_{f}\right\RightPost{par}}\)
is the value of the scale factor at the first
oscillation of the wave after leaving the under-barrier region. It is
seen from Fig. 2 that different waves, that is waves with different wave
numbers
\( n\), stay under the potential barrier for different
intervals of time. This means that, in general, the amplification
coefficient depends on
\( n\):
\(
R{\left\LeftPost{par}n\right\RightPost{par}}=1\)
for all
\( n\) above the top of the potential, and
\(
R{\left\LeftPost{par}n\right\RightPost{par}}{\ifmmode\gg\else$\gg$\fi}1\)
for smaller
\(
n\). The initial spectrum of the waves
\(
h{\left\LeftPost{par}n\right\RightPost{par}}=A
{\left\LeftPost{par}n\right\RightPost{par}}
/a\), defined at
some
\(
\SGMPmathgrk{h}\)
well before the interaction began, transforms into
the final spectrum
\(
h{\left\LeftPost{par}n\right\RightPost{par}}=B
{\left\LeftPost{par}n\right\RightPost{par}}
/a\), defined at some
\(
\SGMPmathgrk{h}\)
well after the interaction completed. The
transformation occurs according to the rule:
\(
B{\left\LeftPost{par}n\right\RightPost{par}}=R
{\left\LeftPost{par}n\right\RightPost{par}}
A{\left\LeftPost{par}n\right\RightPost{par}}\).
This is the essence of the mechanism of the
superadiabatic (parametric) amplification of gravitational waves and, in
fact, of any other fluctuations obeying similar equations.\par

\par The initial amplitudes and spectrum of classical waves can be
arbitrary. It is only important to have a nonzero initial amplitude,
otherwise the final amplitude will also be zero. Now, remaining at the
same classical level, we will imitate the quantum zero-point
fluctuations by assuming that they are classical waves with certain
amplitudes and arbitrary phases. Waves with different frequencies have
different amplitudes, so they form some initial vacuum spectrum. In
order to derive the vacuum spectrum, we neglect the interaction with the
gravitational field and consider, essentially, waves in Minkowski
spacetime. The energy density of gravitational waves scales as
\(
\SGMPmathgrk{e}_{g}={\left\LeftPost{par}c^{4}/G\right\RightPost{par}}
{\left\LeftPost{par}h^{2}/\SGMPmathgrk{l}^{2}\right\RightPost{par}}
\).
For a given wavelength $\lambda$ we want to have
{``}a half of the quantum{''} in each mode, that is we want to
have energy
\(
{{1}\over{2}}\hskip 0.167em {\ifmmode\hbar\else$\hbar$\fi}\SGMPmathgrk{w}
\) in the volume
\( =\SGMPmathgrk{l}^{3}\). It follows from this requirement that the vacuum
amplitude of gravitational waves with the wavelength $\lambda$ is equal to
\(
h{\left\LeftPost{par}\SGMPmathgrk{l}\right\RightPost{par}}=l_{pl}/\SGMPmathgrk{l}
\).
Hence, the initial vacuum spectrum of gravitational
waves, defined at some early epoch
\(
\SGMPmathgrk{h}_{b}\), is
\(
h{\left\LeftPost{par}n\right\RightPost{par}}
{\ifmmode\approx\else$\approx$\fi}nl_{pl}
/a_{b}\), where
\( a_{b}\) is the scale factor at that epoch. This is the
spectrum to be transformed by the interaction with the external
gravitational field. The amplification process makes the number of
quanta in each interacting mode much larger than 1/2. The
renormalization (subtraction of
\(
{{1}\over{2}}{\ifmmode\hbar\else$\hbar$\fi}\SGMPmathgrk{w}\)
) does not practically change the energy of the
amplified waves, while it cancels the initial
\(
{{1}\over{2}}{\ifmmode\hbar\else$\hbar$\fi}\SGMPmathgrk{w}\)
of those (high-frequency) waves that did not interact
with the field.\SGMPnewline
\par

\par \underline{{\protect\normalsize {\bf 5.
Graviton Creation in the Inflationary
Universe}}}\par

\par As an illustration, we will apply the above considerations to the
inflationary (De-Sitter) model. In terms of
\(
\SGMPmathgrk{h}\)
-time the De-Sitter solution (3) has the scale factor
\(
a{\left\LeftPost{par}\SGMPmathgrk{h}\right\RightPost{par}}=-c/H_{0}
\SGMPmathgrk{h}\). (It is convenient to have
\(
\SGMPmathgrk{h}\)
negative and growing from
\(
-\infty \).) We assume that the De-Sitter stage ends at some
\(
\SGMPmathgrk{h}=\SGMPmathgrk{h}_{1}{\ifmmode<\else$<$\fi}0\)
and goes over into the radiation-dominated stage with
\(
a{\left\LeftPost{par}\SGMPmathgrk{h}\right\RightPost{par}}=
{\left\LeftPost{par}\SGMPmathgrk{h}
-2\SGMPmathgrk{h}_{1}\right\RightPost{par}}c/H_{0}\SGMPmathgrk{h}^{%
2}_{1}\).
The relevant potential
\(
U{\left\LeftPost{par}\SGMPmathgrk{h}\right\RightPost{par}}\)
is shown in Fig. 3 by a solid line 1. A wave with the
wave number
\( n\),
\(
{\left\LeftPost{par}n\SGMPmathgrk{h}_{1}\right\RightPost{par}}^{%
2}{\ifmmode\ll\else$\ll$\fi}1\),
enters the potential and ceases to oscillate at some
\(
\SGMPmathgrk{h}_{i}\), when
\(
\SGMPmathgrk{l}{\ifmmode\approx\else$\approx$\fi}c\MthAcnt {a}{\dot }/a\),
that is
\(
2\SGMPmathgrk{p}a/n{\ifmmode\approx\else$\approx$\fi}a^{2}/a'\).
This leads to the
entering condition
\(
n\SGMPmathgrk{h}_{i}{\ifmmode\approx\else$\approx$\fi}1\).
For different
\(
n\)
{`}s this condition is satisfied at different
\(
a{\left\LeftPost{par}\SGMPmathgrk{h}_{i}\right\RightPost{par}}:\hskip 0.167em
a{\left\LeftPost{par}\SGMPmathgrk{h}_{i}\right\RightPost{par}}
{\ifmmode\approx\else$\approx$\fi}c
n/H_{0}\).
The waves leave the potential and start oscillating
again at the radiation-dominated stage when the condition
\(
2\SGMPmathgrk{p}a/n{\ifmmode\approx\else$\approx$\fi}a^{2}/a'\)
is satisfied again. This leads to the exiting condition
\(
n{\left\LeftPost{par}\SGMPmathgrk{h}_{f}-2\SGMPmathgrk{h}_{1}\right\RightPost{par}}
{\ifmmode\approx\else$\approx$\fi}1\)
and
\(
a{\left\LeftPost{par}\SGMPmathgrk{h}_{f}\right\RightPost{par}}
{\ifmmode\approx\else$\approx$\fi}c
/nH_{0}\SGMPmathgrk{h}^{2}_{1}{\ifmmode\approx\else$\approx$\fi}a
{\left\LeftPost{par}
\SGMPmathgrk{h}_{i}\right\RightPost{par}}{\left\LeftPost{par}n
\SGMPmathgrk{h}_{%
1}\right\RightPost{par}}^{-2}\). The amplification coefficient
\(
R{\left\LeftPost{par}n\right\RightPost{par}}={\left\LeftPost{par}n
\SGMPmathgrk{h}_{%
1}\right\RightPost{par}}^{-2}\)
is much larger than 1 for
\(
{\left\LeftPost{par}n\SGMPmathgrk{h}_{1}\right\RightPost{par}}^{%
2}{\ifmmode\ll\else$\ll$\fi}1\)
and scales as
\(
R{\left\LeftPost{par}n\right\RightPost{par}}=n^{-2}\) .\par

\par Now we will see how the vacuum spectrum is transformed. By the time
of entering the barrier, the amplitude
\(
h{\left\LeftPost{par}\SGMPmathgrk{h}_{i}\right\RightPost{par}}
{\ifmmode\approx\else$\approx$\fi}n
{\left\LeftPost{par}l_{pl}/a_{b}\right\RightPost{par}}
{\left\LeftPost{par}a_{b}/a{\left\LeftPost{par}\SGMPmathgrk{h}_{i}
\right\RightPost{par}}\right\RightPost{par}}
{\ifmmode\approx\else$\approx$\fi}l_{pl}H_{%
0}/c\)
is the same for all
\( n\). The exiting amplitude is also independent of
\(
n:\hskip 0.167em h{\left\LeftPost{par}\SGMPmathgrk{h}_{f}\right\RightPost{par}}
=l_{pl}H_{0}/c\),
that is all amplified waves start oscillating with
the same amplitude. The fluctuations which start oscillating (enter the
Hubble radius) with the same amplitude are said to have the
{``}flat{''} Harrison-Zeldovich spectrum. One should keep in mind,
however, that the comparatively shorter waves start oscillating earlier
and their amplitudes decrease more by some fixed (present) time
\(
\SGMPmathgrk{h}_{0}:\hskip 0.167em h{\left\LeftPost{par}\SGMPmathgrk{h}_{%
0}\right\RightPost{par}}{\ifmmode\approx\else$\approx$\fi}h{\left\LeftPost{par}
\SGMPmathgrk{h}_{f
}\right\RightPost{par}}{\left\LeftPost{par}a{\left\LeftPost{par}\SGMPmathgrk{h}_{%
f}\right\RightPost{par}}/a_{o}\right\RightPost{par}}
{\ifmmode\approx\else$\approx$\fi}l_{pl}n/{\left\LeftPost{par}n\SGMPmathgrk{h}_{1}
\right\RightPost{par}}^{%
2}a_{0}\).
As expected, the amplification coefficient
\(
R{\left\LeftPost{par}n\right\RightPost{par}}=n^{-2}\)
transforms the initial vacuum spectrum
\(
h{\left\LeftPost{par}n\right\RightPost{par}}=n\)
into the final spectrum
\(
h{\left\LeftPost{par}n\right\RightPost{par}}=n^{-1}\).\par

\par To put some more detail in this discussion above, one can consider
exact solutions to Eq. (6) at the inflationary
\(
{\left\LeftPost{par}i\right\RightPost{par}}\)
 and radiation-dominated
\(
{\left\LeftPost{par}e\right\RightPost{par}}\)
stages. In the region
\(
\SGMPmathgrk{h}{\ifmmode\leq\else$\leq$\fi}\SGMPmathgrk{h}_{1}\)
the general solution to Eq. (6) has the form
\def\XRefId{}
\begin{equation}\SGMPlab\XRefId\vcenter{\halign{\strut\hfil#\hfil&#\hfil\cr
$\displaystyle{\SGMPmathgrk{m}_{i}=A{\left\LeftPost{sqb}\cos
{\left\LeftPost{par}
n\SGMPmathgrk{h}+\SGMPmathgrk{f}\right\RightPost{par}}-{{1}\over{n\SGMPmathgrk{h}
}}\sin
{\left\LeftPost{par}n\SGMPmathgrk{h}+\SGMPmathgrk{f}\right\RightPost{par}}
\right\RightPost{sqb}}}$\cr
}}\end{equation}

\noindent
where A and
\(
\SGMPmathgrk{f}\)
are arbitrary constants. For
\(
\SGMPmathgrk{h}{\ifmmode\geq\else$\geq$\fi}\SGMPmathgrk{h}_{1}\)
 the general solution has the form
\def\XRefId{}
\begin{equation}
\SGMPlab\XRefId\vcenter{\halign{\strut\hfil#\hfil&#\hfil\cr
$\displaystyle{\SGMPmathgrk{m}_{e}=
\hskip 0.167em B\sin {\left\LeftPost{par}
n\SGMPmathgrk{h}\hskip 0.167em +
\hskip 0.167em \SGMPmathgrk{c}\right\RightPost{par}}
,}$\cr
}}
\end{equation}
where
\(
B\)
and
\(
\SGMPmathgrk{c}\)
 are constants to be determined from the conditions
that
\(
\SGMPmathgrk{m}\)
 and
\(
\SGMPmathgrk{m}'\)
 join continuously at
\(
\SGMPmathgrk{h}=\SGMPmathgrk{h}_{1}\).\par

\par It is convenient to introduce
\(
n\SGMPmathgrk{h}_{1}\hskip 0.167em \equiv \hskip 0.167em x
\).
The joining conditions are:
\def\XRefId{}
\begin{equation}
\SGMPlab\XRefId\vcenter{\halign{\strut\hfil#\hfil&#\hfil\cr
$\displaystyle{}$\hfilneg&$\displaystyle{{}A
{\left\LeftPost{sqb}\cos {\left\LeftPost{par}x+
\SGMPmathgrk{f}\right\RightPost{par}}-
{{1}\over{x}}\sin
{\left\LeftPost{par}x+\SGMPmathgrk{f}
\right\RightPost{par}}\right\RightPost{sqb}}
=B\sin {\left\LeftPost{par}x+\SGMPmathgrk{c}
\right\RightPost{par}}}$\cr
$\displaystyle{}$\hfilneg&$\displaystyle{{}A
{\left\LeftPost{sqb}-{{1}\over{x}}
\cos {\left\LeftPost{par}x+\SGMPmathgrk{f}
\right\RightPost{par}}-{\left\LeftPost{par}
1-{{1}\over{x^{2}}}\right\RightPost{par}}
\sin {\left\LeftPost{par}x+\SGMPmathgrk{f}
\right\RightPost{par}}\right\RightPost{sqb}}
=B\cos {\left\LeftPost{par}x+\SGMPmathgrk{c}
\right\RightPost{par}}}$\cr
}}
\end{equation}

\noindent
Their consequence is
\(
{\left\LeftPost{par}B/A\right\RightPost{par}}^{2}=1+x^{-2
}+{\left\LeftPost{par}x^{-4}-2x^{-2}\right\RightPost{par}}
\SGMPlim{r}\sin \SGMPdolim ^{2}{\left\LeftPost{par}
x+\SGMPmathgrk{f}\right\RightPost{par}}\)
\( -x^{-3}\sin 2{\left\LeftPost{par}
x+\SGMPmathgrk{f}\right\RightPost{par}}\).
One can see that
\(
{\left\LeftPost{par}B/A\right\RightPost{par}}^{2}\)
depends on the initial phase $\phi$. After averaging
over the phase $\phi$ one obtains
\(
\overline{{\left\LeftPost{par}B/A\right\RightPost{par}}^{2}}
=1+1/2x^{4}\).
This expression can be taken as the definition of
the amplification coefficient
\(
R{\left\LeftPost{par}n\right\RightPost{par}}\).
It is seen that
\(
R{\left\LeftPost{par}n\right\RightPost{par}}=1\)
for waves above the potential,
\(
x^{2}{\ifmmode\gg\else$\gg$\fi}1\), and
\def\XRefId{}
\begin{equation}
\SGMPlab\XRefId\vcenter{\halign{\strut\hfil#\hfil&#\hfil\cr
$\displaystyle{R{\left\LeftPost{par}n\right\RightPost{par}}
{\ifmmode\approx\else$\approx$\fi}{{1}\over{%
\Rad{2}\DoRad x^{2}}}}$\cr
}}\end{equation}

\noindent
for the longer waves,
\(
x^{2}{\ifmmode\ll\else$\ll$\fi}1\).
This expression is in full agreement with the
previous qualitative estimates.\par

\par The radiation-dominated era ends at some time
\(
\SGMPmathgrk{h}_{2}\)
and goes over into the matter-dominated era with the
scale factor \newline
\(
a{\left\LeftPost{par}\SGMPmathgrk{h}\right\RightPost{par}}=
c{\left\LeftPost{par}\SGMPmathgrk{h}
+\SGMPmathgrk{h}_{2}-4\SGMPmathgrk{h}_{1}\right\RightPost{par}}^{%
2}/4H_{0}\SGMPmathgrk{h}^{2}_{1}{\left\LeftPost{par}
\SGMPmathgrk{h}_{2}-2\SGMPmathgrk{h}_{1}\right\RightPost{par}}
\). For
\(
\SGMPmathgrk{h}{\ifmmode>\else$>$\fi}\SGMPmathgrk{h}_{2}\)
the potential
\(
U{\left\LeftPost{par}\SGMPmathgrk{h}\right\RightPost{par}}\)
is again non-zero, it is shown by the solid line 2 in
Fig. 3. The waves which satisfy the condition
\(
n{\left\LeftPost{par}\SGMPmathgrk{h}_{2}-
2\SGMPmathgrk{h}_{1}\right\RightPost{par}}
{\ifmmode\gg\else$\gg$\fi}1\)
do not interact with the potential and their
evolution was fully described above. (This part of the spectrum was
first discussed in Ref. 10.) However, the longer waves, satisfying the
opposite condition
\(
n{\left\LeftPost{par}\SGMPmathgrk{h}_{2}-
2\SGMPmathgrk{h}_{1}\right\RightPost{par}}
{\ifmmode\ll\else$\ll$\fi}1\),
interact with the second barrier and transform
additionally their spectrum. (This part of the spectrum was first
discussed in Ref. 11, see also Ref. 12.) These waves start oscillating
at the matter-domianted stage when the requirement
\(
n{\left\LeftPost{par}\SGMPmathgrk{h}_{f}+\SGMPmathgrk{h}_{2}-
4\SGMPmathgrk{h}_{%
1}\right\RightPost{par}}{\ifmmode\approx\else$\approx$\fi}1\)
is satisfired. For these waves, the exiting value of
the scale factor is
\(
a{\left\LeftPost{par}\SGMPmathgrk{h}_{f}\right\RightPost{par}}
{\ifmmode\approx\else$\approx$\fi}c
/H_{0}{\left\LeftPost{par}n\SGMPmathgrk{h}_{1}\right\RightPost{par}}^{%
2}{\left\LeftPost{par}\SGMPmathgrk{h}_{2}-2\SGMPmathgrk{h}_{1}
\right\RightPost{par}}\)
and the amplification coefficient
\(
R{\left\LeftPost{par}n\right\RightPost{par}}
{\ifmmode\approx\else$\approx$\fi}1/{\left\LeftPost{par}n\SGMPmathgrk{h}_{%
1}\right\RightPost{par}}^{2}n{\left\LeftPost{par}\SGMPmathgrk{h}_{%
2}-2\SGMPmathgrk{h}_{1}\right\RightPost{par}}=n^{%
-3}\).
The present-day amplitudes are
\(
h{\left\LeftPost{par}\SGMPmathgrk{h}_{0}\right\RightPost{par}}
{\ifmmode\approx\else$\approx$\fi}l_{%
pl}n/{\left\LeftPost{par}n\SGMPmathgrk{h}_{1}\right\RightPost{par}}^{%
2}n{\left\LeftPost{par}\SGMPmathgrk{h}_{2}-2\SGMPmathgrk{h}_{1}
\right\RightPost{par}}a_{0}=n^{-2}\).\par

\par The waves with the wave number
\( n_{H}\) satisfying
\(
n_{H}{\left\LeftPost{par}\SGMPmathgrk{h}_{0}+\SGMPmathgrk{h}_{%
2}-4\SGMPmathgrk{h}_{1}\right\RightPost{par}}
{\ifmmode\approx\else$\approx$\fi}1\)
enter the Hubble radius at the present epoch. Their wavelengths
\(
\SGMPmathgrk{l}_{H}\)
are of the order of the present Hubble radius,
\(
\SGMPmathgrk{l}_{H}{\ifmmode\approx\else$\approx$\fi}
2\cdot 10^{28}\) cm, and their frequencies
\(
\SGMPmathgrk{n}_{H}\)
are
\(
\SGMPmathgrk{n}_{H}{\ifmmode\approx\else$\approx$\fi}10^{-18}\) Hz.
The amplitude of these waves is, roughly,
\(
h_{H}{\left\LeftPost{par}\SGMPmathgrk{h}_{0}\right\RightPost{par}}
{\ifmmode\approx\else$\approx$\fi}l_{pl}H_{0}/c\).
It can not be too large in order not to cause too
large angular (quadrupole) variations in the temperature of the
microwave background radiation,
\(
\bigtriangleup T/T{\ifmmode\approx\else$\approx$\fi}h_{H}{\left\LeftPost{par}
\SGMPmathgrk{h}_{0}
\right\RightPost{par}}\).
This requirement places a limit on the value of the
Hubble parameter
\(
H_{0}\)
at the De-Sitter stage [11]. Equally strong limit
follows from the observational restrictions on
\(
\bigtriangleup T/T\)
in a few degrees angular scale, where the
contribution of waves with frequencies
\(
\SGMPmathgrk{n}=10^{-16}\)
\(
Hz\)
is dominant [9] . The inflationary spectrum of
graviational waves in terms of the spectral flux density as a function
of frequency
\(
\SGMPmathgrk{n}\), is shown in Fig. 4 (adopted from [9]). The upper
position of the spectrum is determined by the observational data on
\(
\bigtriangleup T/T\).
At the same graph one can see theoretical
predictions of some other models and the existing experimental limits as
well as the expected levels of sensitivity of various observational
techniques.\par

\par The predicted gravity-wave spectrum is more complex if the Hubble
parameter at the inflationary stage was not constant. It is interesting
to know that the time variations of the Hubble parameter are in
one-to-one correspondence with the frequency variations of the
present-day spectral energy density of waves [13]. This makes it
possible, at least in principle, to study the details of the very early
evolution of the universe by measuring the spectral properties of relic
gravitational waves.\par

\par In our previous discussion we have been mainly interested in the
amplitude of a {``}typical{''} wave. In other words, we have been
calculating the r.m.s. value of the final amplitude, assuming that the
initial phase
\(
\SGMPmathgrk{f}\)
is distributed randomly and evenly in the interval
form 0 to
\(
2\SGMPmathgrk{p}\).
However, the distribution of the final phase is also
important. Equivalently, one can ask about the final distributions of
the quadrature components of the wave, that is the components
proportional to
\(
\sin n\SGMPmathgrk{h}\)
and
\(
\cos n\SGMPmathgrk{h}\).
This study can serve as an introduction to the
notion of the quantum mechanical squeezing which we will be discussing
later.\par

\par Let us return to the exact solutions (7) and (8). Initially, for
\(
\SGMPmathgrk{h}{\ifmmode\rightarrow\else$\rightarrow$\fi}-\infty \),
solution (7) can be written as
\(
\SGMPmathgrk{m}_{i}{\ifmmode\approx\else$\approx$\fi}A
{\left\LeftPost{sqb}v_{1}\hskip 0.167em
\sin n\SGMPmathgrk{h}+v_{2}\cos n\SGMPmathgrk{h}\right\RightPost{sqb}}
\),
where
\(
v_{1}\equiv -\sin \SGMPmathgrk{f},\hskip 0.167em v_{%
2}\equiv \cos \SGMPmathgrk{f}\).
The mean values of
\(
v_{1},\hskip 0.167em v_{2}\)
and
\(
v_{1}v_{2}\)
are zero. However, the mean values of
\(
v^{2}_{1}\)
and
\(
v^{2}_{2}\)
are nonzero and equal:
\(
\overline{v^{2}_{1}}=\overline{v^{2}_{2
}}={{1}\over{2}}\).
To derive the quadrature components of solution (8)
one can first find, from the joining conditions, the constant
\(
\SGMPmathgrk{c}\)
and rewrite Eq. (8) in the form:
\def\XRefId{}
\begin{equation}\SGMPlab\XRefId\vcenter{\halign{\strut\hfil#\hfil&#\hfil\cr
$\displaystyle{\SGMPmathgrk{m}_{e}=A\sin n\SGMPmathgrk{h}{\left\LeftPost{sqb}
+{{1}\over{x^{2}}}\cos x\sin
{\left\LeftPost{par}x+\SGMPmathgrk{f}\right\RightPost{par}}-{{1\hskip -0.167em
}\over{x}}\cos \SGMPmathgrk{f}-\sin \SGMPmathgrk{f}
\right\RightPost{sqb}}+}$\cr
$\displaystyle{A\cos n\SGMPmathgrk{h}{\left\LeftPost{sqb}-{{1}\over{x^{%
2}}}\sin x\sin {\left\LeftPost{par}x+\SGMPmathgrk{f}
\right\RightPost{par}}-{{1}\over{x}}\sin
\SGMPmathgrk{f}+\cos \SGMPmathgrk{f}\right\RightPost{sqb}}\equiv }$\cr
$\displaystyle{Ak_{1}\sin n\SGMPmathgrk{h}+Ak_{2}\cos
n\SGMPmathgrk{h}}$\cr
}}\end{equation}

\noindent
The mean values of
\(
k_{1}\)
and
\(
k_{2}\)
are again zero, but for the quadratic combinations
one obtains:
\def\XRefId{}
\begin{equation}\SGMPlab\XRefId\vcenter{\halign{\strut\hfil#\hfil&#\hfil\cr
$\displaystyle{{}\overline{k^{2}_{1}}={{%
1}\over{2}}{\left\LeftPost{sqb}1+{{1}\over{x^{%
2}}}+{{1}\over{x^{2}}}{\left\LeftPost{par}
{{1}\over{x^{2}}}-2\right\RightPost{par}}
\SGMPlim{r}\cos \SGMPdolim ^{2}x-{{%
1}\over{x^{3}}}\sin 2x\right\RightPost{sqb}}
,}$\cr
$\displaystyle{{}\overline{k^{2}_{2}}={{%
1}\over{2}}{\left\LeftPost{sqb}1+{{1}\over{x^{%
2}}}+{{1}\over{x^{2}}}{\left\LeftPost{par}
{{1}\over{x^{2}}}-2\right\RightPost{par}}
\SGMPlim{r}\sin \SGMPdolim ^{2}x+{{%
1}\over{x^{3}}}\sin 2x\right\RightPost{sqb}}
,}$\cr
$\displaystyle{\overline{k_{1}k_{2}}={{1}\over{2
}}{\left\LeftPost{sqb}-{{1}\over{2x^{2}}}
{\left\LeftPost{par}{{1}\over{x^{2}}}-2\right\RightPost{par}}
\sin 2x-{{1}\over{x^{3}}}\cos
2x\right\RightPost{sqb}}}$\cr
}}\end{equation}

Note that
\(
\overline{k^{2}_{1}}\hskip 0.167em \overline{k^{%
2}_{2}}-\overline{{\left\LeftPost{par}k_{1}k_{%
2}\right\RightPost{par}}^{2}}={{1}
\over{4}}=\overline{v^{2}_{1}}\hskip 0.167em
\overline{v^{2}_{2}}\).
If one makes a constant shift
\(
\SGMPmathgrk{h}=\MthAcnt {\SGMPmathgrk{h}}{\tilde }+y\),
the solution (11) transforms to
\(
\SGMPmathgrk{m}_{e}=Al_{1}\sin n\MthAcnt {\SGMPmathgrk{h}
}{\tilde }+Al_{2}\cos n\MthAcnt {\SGMPmathgrk{h}
}{\tilde }\).
The constant
\(
y\)
 can be chosen in such a way that
\(
\overline{l_{1}l_{2}}=0\)
. Under this choice one obtains
\def\XRefId{}
\begin{equation}\SGMPlab\XRefId\vcenter{\halign{\strut\hfil#\hfil&#\hfil\cr
$\displaystyle{\overline{l^{2}_{1}}={{1}\over{4}}
{\left\LeftPost{sqb}2+{{1}\over{x^{4}}}+{{%
1}\over{x^{2}}}\Rad{4+{{1}\over{%
x^{4}}}}\DoRad \right\RightPost{sqb}},\hskip 0.265em
\hskip 0.265em \hskip 0.265em \overline{l^{2}_{2}
}={{1}\over{4}}{\left\LeftPost{sqb}2+{{1
}\over{x^{4}}}-{{1}\over{x^{2}
}}\Rad{4+{{1}\over{x^{4}}}
}\DoRad \right\RightPost{sqb}}}$\cr
}}\end{equation}

\noindent
This choice minimizes one of the quadrature variances and maximizes the
other.\par

\par We are interested in the case
\(
x^{2}{\ifmmode\ll\else$\ll$\fi}1\). For this case,
\(
\overline{l^{2}_{1}}{\ifmmode\approx\else$\approx$\fi}1/2x^{4},\hskip 0.167em
\overline{l^{2}_{2}}{\ifmmode\approx\else$\approx$\fi}x^{4}/2\).
We see that during the amplification process, one of
the noise components strongly increased while the other decreased
equally strongly. One can also say that the final phase
\( \SGMPmathgrk{c}\)
is not evenly distributed as the function of
\( \SGMPmathgrk{f}\),
but is highly peaked near the values
\(
tg\hskip 0.265em \hskip 0.265em \SGMPmathgrk{c}
{\ifmmode\approx\else$\approx$\fi}-2x\).
We will see below that the supression (squeezing) of
variances in one of two quadrature components of the wave field is a
characteristic feature of squeezed quantum states. Moreover, squeezing
may reduce one of the two variances below the level of zero-point
quantum fluctuations.\SGMPnewline
\par

\par \underline{{\protect\normalsize {\bf 6.
Quantum States of a Harmonic
Oscillator}}}\par

\par Let us first recall some properties of quantum states of an
ordinary harmonic oscillator. We will need this information in our
further discussion. Especially, we will be interested in the notion of
squeezed quantum states.\par

\par Classical equations of motion for a harmonic oscillator,
\(
\MthAcnt {x}{\ddot }+\SGMPmathgrk{w}^{2}x=0\),
can be derived from the Lagrange function
\(
L=\hskip 0.167em {{1}\over{2}}m\MthAcnt {x^{%
2}}{\dot }-{{1}\over{2}}mw^{%
2}x^{2}\)
according to the rule:
\(
{\left\LeftPost{par}{{\SGMPmathgrk{6}L}\over{\SGMPmathgrk{6}\MthAcnt {x}{\dot }
}}\right\RightPost{par}}^{.}\hskip 0.167em \hskip 0.167em
-\hskip 0.167em {{\SGMPmathgrk{6}L}\over{\SGMPmathgrk{6}x}}
=0\).
Associated with L is the Hamilton function
\(
H=\hskip 0.167em {{p^{2}}\over{2m}}\hskip 0.167em
+\hskip 0.167em {{m}\over{2}}w^{2}x^{%
2}\),
where
\(
p={\left\LeftPost{par}{{\SGMPmathgrk{6}L}\over{\SGMPmathgrk{6}\MthAcnt {x}{\dot
}
}}\right\RightPost{par}}\).
Quantization is achieved by introducing the operators
\( \MthAcnt {x}{\hat }\)
and
\(
\MthAcnt {p}{\hat }\hskip 0.167em =\hskip 0.167em
-\hskip 0.167em i{\ifmmode\hbar\else$\hbar$\fi}
{{\SGMPmathgrk{6}}\over{\SGMPmathgrk{6}x}}
\)
and establishing the commutation relation:
\(
{\left\LeftPost{sqb}\MthAcnt {x}{\hat }\MthAcnt {p}{\hat }
\right\RightPost{sqb}}\hskip 0.167em =\hskip 0.167em
i{\ifmmode\hbar\else$\hbar$\fi}
.\)
 From
\(
\MthAcnt {x}{\hat }\)
and
\(
\MthAcnt {p}{\hat }\)
one can construct the creation and annihilation
operators a${}^{+}$ and a:
\(
a^{+}\hskip 0.167em =\hskip 0.167em {\left\LeftPost{par}{{%
mw}\over{2{\ifmmode\hbar\else$\hbar$\fi}}}\right\RightPost{par}}^{1/2}
{\left\LeftPost{par}\MthAcnt {x}{\hat }\hskip 0.167em -
\hskip 0.167em i{{\MthAcnt {p}{\hat }}\over{%
mw}}\right\RightPost{par}},\hskip 0.265em a={\left\LeftPost{par}
{{mw}\over{2{\ifmmode\hbar\else$\hbar$\fi}}}\right\RightPost{par}}^{%
1/2}{\left\LeftPost{par}\MthAcnt {x}{\hat }+i{{%
\MthAcnt {p}{\hat }}\over{mw}}\right\RightPost{par}}
,\hskip 0.265em {\left\LeftPost{sqb}a,a^{+}\right\RightPost{sqb}}
=1\)
and the particle number operator
\(
\MthAcnt {N}{\hat }\):
\(
\MthAcnt {N}{\hat }=\MthAcnt {a}{\hat }^{%
+}\MthAcnt {a}{\hat }={{\MthAcnt {H}{\hat }
}\over{{\ifmmode\hbar\else$\hbar$\fi}w}}-{{1}\over{2}}\).
The oscillator can be described by the wave function
(or state function)
\(
\SGMPmathgrk{y}{\left\LeftPost{par}x,t\right\RightPost{par}}\)
which satisfy the Schr\"odinger equation:
\(
i{\ifmmode\hbar\else$\hbar$\fi}
{{\SGMPmathgrk{6}\SGMPmathgrk{y}}\over{\SGMPmathgrk{6}t}}
=\MthAcnt {H}{\hat }\SGMPmathgrk{y}\).
The ground (vacuum) quantum state
{\ifmmode|\else$|$\fi}0{\ifmmode>\else$>$\fi} is defined by
the requirement a{\ifmmode|\else$|$\fi}0{\ifmmode>\else$>$\fi}=0.
The ground wave function has the form
\(
\SGMPmathgrk{y}{\left\LeftPost{par}x\right\RightPost{par}}=
{\left\LeftPost{par}{{%
mw}\over{\SGMPmathgrk{p}{\ifmmode\hbar\else$\hbar$\fi}}}
\right\RightPost{par}}^{%
1/4}e^{-{{mw}\over{2{\ifmmode\hbar\else$\hbar$\fi}}}x^{2
}}\).
The n-quantum states are defined as the eigenstates of the
\( \MthAcnt {N}{\hat }\) operator:
\(
\MthAcnt {N}{\hat }{\left\LeftPost{vb}n{\ifmmode>\else$>$\fi}
\hskip 0.167em
=n\right\RightPost{vb}}n{\ifmmode>\else$>$\fi},\)
they are also eigenstates of
\(
\MthAcnt {H}{\hat }\)
with eigenvalues
\(
{\ifmmode\hbar\else$\hbar$\fi}\SGMPmathgrk{w}{\left\LeftPost{par}n+
{{1}\over{2}}
\right\RightPost{par}},\hskip 0.265em \MthAcnt {H}{\hat }
{\left\LeftPost{vb}n{\ifmmode>\else$>$\fi}
={\ifmmode\hbar\else$\hbar$\fi}w{\left\LeftPost{par}n+{{1}\over{%
2}}\right\RightPost{par}}\right\RightPost{vb}}n
{\ifmmode>\else$>$\fi}\).
These states are produced by the action of the
creation operator a${}^{+}$ on the vacuum state:
\(
{\ifmmode |\else$ |$\fi}n{\ifmmode>\else$>$\fi}={{{\left\LeftPost{par}
a^{+}\right\RightPost{par}}^{%
n}}\over{\Rad{n!}\DoRad }}{\ifmmode |\else$ |$\fi}0{\ifmmode>\else$>$\fi}\)
.\par

\par An important class of quantum states are coherent states which are
regarded as the {``}most classical{''}. The coherent states are
generated from the vacuum state {\ifmmode|\else$|$\fi}0{\ifmmode>\else$>$\fi}
by the action of the displacement
operator:
\(
D{\left\LeftPost{par}a,\SGMPmathgrk{a}\right\RightPost{par}}\equiv \hskip
0.167em
\exp {\left\LeftPost{sqb}\SGMPmathgrk{a}a^{+}-\SGMPmathgrk{a}^{%
*}a\right\RightPost{sqb}}\),
where $\alpha$ is an arbitrary complex number. We are
mostly interested in squeezed states. Squeezed states involve operators
quadratic in
\( a,\hskip 0.167em a^{+}\).
A one-mode squeezed state (for a review see, for
example, [13]) is generated by the action of the squeeze operator:\newline
\(
S_{1}{\left\LeftPost{par}r,\SGMPmathgrk{f}\right\RightPost{par}}\equiv
exp{\left\LeftPost{sqb}{{1}\over{2}}r{\left\LeftPost{par}
e^{-2i\SGMPmathgrk{f}}a^{2}-e^{2i\SGMPmathgrk{f}}a^{%
+2}\right\RightPost{par}}\right\RightPost{sqb}}\),
where the real numbers
\( r\) and
\( \SGMPmathgrk{f}\)
are known as the squeeze factor and squeeze angle:
\(
0{\ifmmode\leq\else$\leq$\fi}r{\ifmmode<\else$<$\fi}\infty ,
\hskip 0.265em \hskip 0.265em -{{\SGMPmathgrk{p}
}\over{2}}{\ifmmode<\else$<$\fi}\SGMPmathgrk{f}
{\ifmmode\leq\else$\leq$\fi}{{\SGMPmathgrk{p}}\over{%
2}}\).
A squeezed state can be generated by the action of
the squeeze operator on any coherent state and, in particular, on the
vacuum state, in which case it is called squeezed vacuum state. If a
harmonic oscillator is exposed to the time-dependent interaction, a
one-mode squeezed state is produced, as a result of the Schr\"odinger
evolution, by the interaction Hamiltonian
\def\XRefId{}
\begin{equation}\SGMPlab\XRefId\vcenter{\halign{\strut\hfil#\hfil&#\hfil\cr
$\displaystyle{H{\left\LeftPost{par}t\right\RightPost{par}}=
\SGMPmathgrk{s}{\left\LeftPost{par}
t\right\RightPost{par}}a^{+2}+\SGMPmathgrk{s}^{*}{\left\LeftPost{par}
t\right\RightPost{par}}a^{2}}$\cr
}}\end{equation}

\noindent
where
\( \SGMPmathgrk{s}\)
is arbitrary function of time.\par

\par The meaning of the word {``}squeezing {''} is related to the
properties of these states with respect to variances (or noise moments)
$\Delta$A of different operators A:
\(
\SGMPmathgrk{D}A\equiv A-{\ifmmode<\else$<$\fi}A{\ifmmode>\else$>$\fi}\).
The squeezed wave functions are always Gaussian:
\def\XRefId{}
\begin{equation}\SGMPlab\XRefId\vcenter{\halign{\strut\hfil#\hfil&#\hfil\cr
$\displaystyle{\SGMPmathgrk{y}{\left\LeftPost{par}x\right\RightPost{par}}
\sim e^{-{{%
1}\over{2}}\SGMPmathgrk{g}x^{2}}}$\cr
}}\end{equation}
but, the variances of the variables
\(
\MthAcnt {x}{\hat }\)
and
\(
\MthAcnt {p}{\hat }\)
are substantially different. They can be presented in
terms of the complex parameter
\(
\SGMPmathgrk{g},\SGMPmathgrk{g}\equiv \SGMPmathgrk{g}_{1}+i\SGMPmathgrk{g}_{2}
,\)
or real parameters r, $\phi$:
\def\XRefId{}
\begin{equation}\SGMPlab\XRefId\vcenter{\halign{\strut\hfil#\hfil&#\hfil\cr
$\displaystyle{}$\hfilneg&$\displaystyle{{}{\ifmmode<\else$<$\fi}
{\left\LeftPost{par}\SGMPmathgrk{D}\MthAcnt {x}{\hat }
\right\RightPost{par}}^{2}{\ifmmode>\else$>$\fi}={{1}\over{2\SGMPmathgrk{g}_{%
1}}},\hskip 0.265em \hskip 0.265em \hskip 0.265em
{\ifmmode<\else$<$\fi}{\left\LeftPost{par}\SGMPmathgrk{D}
\MthAcnt {p}{\hat }\right\RightPost{par}}^{%
2}{\ifmmode>\else$>$\fi}={{{\left\LeftPost{vb}
\SGMPmathgrk{g}\right\RightPost{vb}}^{%
2}}\over{2\SGMPmathgrk{g}_{1}}},}$\cr
$\displaystyle{}$\hfilneg&$\displaystyle{{}
{\ifmmode<\else$<$\fi}{\left\LeftPost{par}\SGMPmathgrk{D}\MthAcnt {x}{\hat }^{%
2}\right\RightPost{par}}{\ifmmode>\else$>$\fi}={{1}\over{2}}
{\left\LeftPost{par}ch2r-sh2r\cos 2\SGMPmathgrk{f}\right\RightPost{par}}
,}$\cr
$\displaystyle{}$\hfilneg&$\displaystyle{{}
{\ifmmode<\else$<$\fi}{\left\LeftPost{par}
\SGMPmathgrk{D}\MthAcnt {p}{\hat }\right\RightPost{par}}^{%
2}{\ifmmode>\else$>$\fi}={{1}\over{2}}{\left\LeftPost{par}
ch2r+sh2r\cos 2\SGMPmathgrk{f}\right\RightPost{par}}}$\cr
}}\end{equation}

\par

\par These variances should be compared with those for a coherent state,
in which case they are always equal to each other and are minimally
possible:
\(
{\ifmmode<\else$<$\fi}{\left\LeftPost{par}\SGMPmathgrk{D}
\MthAcnt {x}{\hat }\right\RightPost{par}}^{%
2}{\ifmmode>\else$>$\fi}={\ifmmode<\else$<$\fi}{\left\LeftPost{par}
\SGMPmathgrk{D}\MthAcnt {p}{\hat }
\right\RightPost{par}}^{2}{\ifmmode>\else$>$\fi}={{1}\over{2}}
\).
So, in a squeezed state, one component of the noise
is always large but another is {``}squeezed{''} and can be smaller
than
\(
{{1}\over{2}}\).
In (x,p) plane the line of a total noise
\(
K={{1}\over{2}}{\left\LeftPost{sqb}{\ifmmode<\else$<$\fi}{\left\LeftPost{par}
\SGMPmathgrk{D}\MthAcnt {x}{\hat }\right\RightPost{par}}^{%
2}{\ifmmode>\else$>$\fi}+{\ifmmode<\else$<$\fi}{\left\LeftPost{par}
\SGMPmathgrk{D}\MthAcnt {p}{\hat }
\right\RightPost{par}}^{2}{\ifmmode>\else$>$\fi}\right\RightPost{sqb}}\)
for the coherent states can be described by a circle,
while this line is an ellipse for the squeezed states (see Fig. 5). In a
squeezed vacuum state the mean values of
\(
\MthAcnt {x}{\hat }\)
and
\(
\MthAcnt {p}{\hat }\)
are zero, so in this case, the center of the ellipse
is at the origin.\par

\par The mean value of the particle number operator
\( N\)
is not zero in a squeezed vacuum state, it can be
expressed in terms of the squeeze parameter
\(
r:\hskip 0.265em \hskip 0.265em \hskip 0.265em
{\ifmmode<\else$<$\fi}N{\ifmmode>\else$>$\fi}=s
h^{2}r\).
For strongly squezed states,
\(
r{\ifmmode\gg\else$\gg$\fi}1\), the
\(
{\ifmmode<\else$<$\fi}N{\ifmmode>\else$>$\fi}\)
is very large
\(
{\ifmmode<\else$<$\fi}N{\ifmmode>\else$>$\fi}
{\ifmmode\approx\else$\approx$\fi}{{1}\over{4}}e^{2r}\).
The variance of
\(
N\)
is
\(
{\ifmmode<\else$<$\fi}{\left\LeftPost{par}\bigtriangleup
N\right\RightPost{par}}^{2}
{\ifmmode>\else$>$\fi}={{1}\over{2}}sh^{2}2r\). The
\(
{\ifmmode<\else$<$\fi}{\left\LeftPost{par}\bigtriangleup
N\right\RightPost{par}}^{2}
{\ifmmode>\else$>$\fi}\)
is also very large for
\(
r{\ifmmode\gg\else$\gg$\fi}1:\hskip 0.265em \hskip 0.265em \hskip 0.265em
{\ifmmode<\else$<$\fi}{\left\LeftPost{par}
\bigtriangleup N\right\RightPost{par}}^{2}{\ifmmode>\else$>$\fi}
{\ifmmode\approx\else$\approx$\fi}{{1
}\over{8}}e^{4r}\)
and
\(
{\ifmmode<\else$<$\fi}{\left\LeftPost{par}\bigtriangleup
N\right\RightPost{par}}^{2}
{\ifmmode>\else$>$\fi}^{{{1}\over{2}}}{\ifmmode\approx\else$\approx$\fi}{{1}
\over{2\Rad{2}\DoRad }}e^{2r}\),
that is
\(
{\ifmmode<\else$<$\fi}
{\left\LeftPost{par}\bigtriangleup N\right\RightPost{par}}^{2}
{\ifmmode>\else$>$\fi}^{{{1}\over{2}}}
{\ifmmode\approx\else$\approx$\fi}{\ifmmode<\else$<$\fi}N{\ifmmode>\else$>$\fi}
\).
In contrast,
\(
{\ifmmode<\else$<$\fi}{\left\LeftPost{par}\bigtriangleup
N\right\RightPost{par}}^{2}
{\ifmmode>\else$>$\fi}^{{{1}\over{2}}}\hskip 0.265em \hskip 0.265em
=\hskip 0.265em \hskip 0.265em
{\ifmmode<\else$<$\fi}N{\ifmmode>\else$>$\fi}^{{{1}
\over{2}}}\)
for coherent states.\par

\par A notion which is very useful for our problem of the pair particle
creation is the two-mode squeezed states. The two modes under our
consideration will be two particles (waves) travelling in the opposite
direction. A two-mode squeezed vacuum state is generated from the vacuum
\(
{\ifmmode |\else$ |$\fi}0,0{\ifmmode>\else$>$\fi}\)
by the action of the two-mode squeeze operator
\def\XRefId{}
\begin{equation}\SGMPlab\XRefId\vcenter{\halign{\strut\hfil#\hfil&#\hfil\cr
$\displaystyle{S{\left\LeftPost{par}r,
\hskip 0.167em \SGMPmathgrk{f}\right\RightPost{par}}
={\rm exp}{\left\LeftPost{sqb}r{\left\LeftPost{par}
e^{-2i\SGMPmathgrk{f}}
a_{+}a_{-}-e^{2i\SGMPmathgrk{f}}a^{+}_{%
+}a^{+}_{-}\right\RightPost{par}}\right\RightPost{sqb}}
}$\cr
}}\end{equation}
where
\(
a_{+},\hskip 0.167em a_{-}\)
and
\(
a^{+}_{+},\hskip 0.167em a^{+}_{-
}\)
are annihilation and creation operators for the two
modes,
\(
r\)
is the squeeze parameter and
\(
\SGMPmathgrk{f}\)
is the squeeze angle.\SGMPnewline
\par

\par \underline{{\protect\normalsize {\bf 7.
Squeezed quantum states of relic gravitons and }}}\newline
\underline {{{\bf primordial density
pertrubations}}}\par

\par Our preceding analyses of perturbations interacting with the
variable gravitational field was essentially classical. Quantum
mechanics entered our calculations only as a motivation for choosing the
particular initial amplitudes and for making the averaging over initial
phases. We interpreted the final results as quantum-mechanical
generation of gravitational waves and, possibly, other perturbations,
but a rigorous treatment is still needed. We will see now that a
consistent quantum-mechanical theory confirms our main results and gives
a much more detailed and informative picture of the entire
phenomenon.\par

\par We will start again from gravitational waves. The gravitational
wave field
\(
h_{ij}{\left\LeftPost{par}\SGMPmathgrk{h},\hskip 0.167em {\bf x}
\right\RightPost{par}}\)
becomes an operator and can be written in the general form
\def\XRefId{}
\begin{equation}\SGMPlab\XRefId\vcenter{\halign{\strut\hfil#\hfil&#\hfil\cr
$\displaystyle{h_{ij}{\left\LeftPost{par}\SGMPmathgrk{h},\hskip 0.167em {\bf x
}\right\RightPost{par}}=C\int _{-\infty }^{%
\infty }d^{3}{\bf n}\sum _{s=1}
^{2}p^{s}_{ij}{\left\LeftPost{par}{\bf n}
\right\RightPost{par}}{\left\LeftPost{sqb}a^{s}_{{\bf n}
}{\left\LeftPost{par}\SGMPmathgrk{h}\right\RightPost{par}}e^{i{\bf nx
}}+a^{s+}_{{\bf n}}{\left\LeftPost{par}\SGMPmathgrk{h}
\right\RightPost{par}}e^{-i{\bf nx}}\right\RightPost{sqb}}
}$\cr
}}\end{equation}

\noindent
This expression requires some explanation. In Eq. (18) we do not write
the scale factor
\(
a{\left\LeftPost{par}\SGMPmathgrk{h}\right\RightPost{par}}\)
in front of the expression (compare with Eq. (5))
which can be taken care of later. It is precisely
\(
h_{ij}\)
{`}s given by Eq. (18) that appear automatically in the
{``}field-theoretical{''} treatment of the problem, see Refs. [15,
16]. The normalization constant
\( C\)
includes all the numerical coefficients but we do not
need them now and will not write
\( C\)
below. We will also use units
\(
c=1,\hskip 0.167em {\ifmmode\hbar\else$\hbar$\fi}=1\).
Two tensors
\(
p^{s}_{ij}{\left\LeftPost{par}{\bf n}\right\RightPost{par}},\hskip 0.167em
s=1,\hskip 0.167em 2\),
represent two independent polarization states of each mode (wave). The tensors
\( p^{s}_{ij}\)
satisfy the {``}transverse-traceless{''}
conditions:
\(
p^{s}_{ij}n^{j}=0,\hskip 0.167em p^{%
s}_{ij}\SGMPmathgrk{d}^{ij}=0\).
For a wave travelling in the firection
\(
{{{\bf n}}\over{n}}={\left\LeftPost{par}\sin
\SGMPmathgrk{q}\cos \SGMPmathgrk{4},\hskip 0.167em \sin \SGMPmathgrk{q}
\sin \SGMPmathgrk{4},\hskip 0.167em \cos \SGMPmathgrk{q}\right\RightPost{par}}
\)
the two polarization tensors are
\(
p^{1}_{ij}{\left\LeftPost{par}{\bf n}\right\RightPost{par}}
=l_{i}l_{j}-m_{i}m_{j},\hskip 0.167em
p^{2}_{ij}{\left\LeftPost{par}{\bf n}\right\RightPost{par}}
=l_{i}m_{j}+l_{j}m_{i}\)
where
\(
l_{j},\hskip 0.167em m_{j}\)
are two unit vectors ortogonal to
\(
{{{\bf n}}\over{n}}\)
and to each other:
\(
l_{j}={\left\LeftPost{par}\sin \SGMPmathgrk{4},\hskip 0.167em
-\cos \SGMPmathgrk{4},\hskip 0.167em 0\right\RightPost{par}},
\hskip 0.167em m_{j}={\left\LeftPost{par}\cos \SGMPmathgrk{q}
\cos \SGMPmathgrk{4},\hskip 0.167em \cos \SGMPmathgrk{q}\sin
\SGMPmathgrk{4},\hskip 0.167em -\sin \SGMPmathgrk{q}\right\RightPost{par}}
\).
The operators
\(
a^{s}_{{\bf n}}{\left\LeftPost{par}\SGMPmathgrk{h}\right\RightPost{par}}
,\hskip 0.167em a^{s+}_{{\bf n}}{\left\LeftPost{par}
\SGMPmathgrk{h}\right\RightPost{par}}\)
are annihilation and creation operators for waves
(particles) travelling in the direction of
\( {\bf n}\).
These are Heisenberg operators depending on time
\(
\SGMPmathgrk{h}\).\par

\par The evolution of the operators
\(
a^{s}_{{\bf n}}{\left\LeftPost{par}\SGMPmathgrk{h}\right\RightPost{par}}
,\hskip 0.167em a^{s+}_{{\bf n}}{\left\LeftPost{par}
\SGMPmathgrk{h}\right\RightPost{par}}\)
is governed by the Heisenberg equations of motion for each mode
\( {\bf n}\) and for each polarization
\( s\):
\def\XRefId{}
\begin{equation}\SGMPlab\XRefId\vcenter{\halign{\strut\hfil#\hfil&#\hfil\cr
$\displaystyle{{{da_{{\bf n}}}\over{d\SGMPmathgrk{h}}}
=-i{\left\LeftPost{sqb}a_{{\bf n}},\hskip 0.167em H\right\RightPost{sqb}}
,\hskip 0.265em \hskip 0.265em \hskip 0.265em \hskip 0.167em
{{da^{+}_{{\bf n}}}\over{d\SGMPmathgrk{h}}}
=-i{\left\LeftPost{sqb}a^{+}_{{\bf n}},\hskip 0.167em
H\right\RightPost{sqb}}}$\cr
}}\end{equation}

\noindent
where
\( H\) is the Hamiltonian of the problem. To derive the
Hamiltonian we will proceed as follows.\par

\par The classical equations (6) are the Euler-Lagrange equations
following from the Lagrangian
\def\XRefId{}
\begin{equation}\SGMPlab\XRefId\vcenter{\halign{\strut\hfil#\hfil&#\hfil\cr
$\displaystyle{L={{1}\over{2}}{\left\LeftPost{sqb}\SGMPmathgrk{m}'^{%
2}-n^{2}\SGMPmathgrk{m}^{2}-2{{a'}\over{%
a}}\SGMPmathgrk{m}\SGMPmathgrk{m}'+{\left\LeftPost{par}{{a'}\over{%
a}}\right\RightPost{par}}^{2}\SGMPmathgrk{m}^{2}
\right\RightPost{sqb}}}$\cr
}}\end{equation}

\noindent
The associated Hamiltonian is
\def\XRefId{}
\begin{equation}\SGMPlab\XRefId\vcenter{\halign{\strut\hfil#\hfil&#\hfil\cr
$\displaystyle{H={{1}\over{2}}{\left\LeftPost{sqb}p^{2}
+n^{2}\SGMPmathgrk{m}^{2}+2{{a'}\over{a}}
\SGMPmathgrk{m}p\right\RightPost{sqb}}}$\cr
}}\end{equation}

\noindent
where
\( p\) is the canonically conjugated momentum:
\(
p=\SGMPmathgrk{6}L/\SGMPmathgrk{6}\SGMPmathgrk{m}'=
\SGMPmathgrk{m}'-{\left\LeftPost{par}a'/a\right\RightPost{par}}
\SGMPmathgrk{m}\).
In the quantum treatment,
\(
\SGMPmathgrk{m}\)
and
\( p\) are operators satisfying the commutation relation
\(
{\left\LeftPost{sqb}\SGMPmathgrk{m},\hskip 0.167em p\right\RightPost{sqb}}
=i\).
The associated annihilation and creation operators are
\def\XRefId{}
\begin{equation}\SGMPlab\XRefId\vcenter{\halign{\strut\hfil#\hfil&#\hfil\cr
$\displaystyle{b=\Rad{{{n}\over{2}}}\DoRad {\left\LeftPost{par}
\SGMPmathgrk{m}+i{{p}\over{n}}\right\RightPost{par}},
\hskip 0.265em \hskip 0.265em \hskip 0.265em b^{+}
=\Rad{{{n}\over{2}}}\DoRad {\left\LeftPost{par}
\SGMPmathgrk{m}-i{{p}\over{n}}\right\RightPost{par}},
\hskip 0.265em \hskip 0.265em \hskip 0.265em {\left\LeftPost{sqb}
b,b^{+}\hskip 0.167em \right\RightPost{sqb}}=1}$\cr
}}\end{equation}

In terms of
\( b,\hskip 0.167em b^{+}\),
the Hamiltonian (21) acquires the form
\def\XRefId{}
\begin{equation}\SGMPlab\XRefId\vcenter{\halign{\strut\hfil#\hfil&#\hfil\cr
$\displaystyle{H=nb^{+}b+\SGMPmathgrk{s}
{\left\LeftPost{par}\SGMPmathgrk{h}\right\RightPost{par}}
b^{+2}+\SGMPmathgrk{s}^{*}{\left\LeftPost{par}\SGMPmathgrk{h}\right\RightPost{par}}
b^{2}}$\cr
}}\end{equation}

\noindent
where the coupling function
\(
\SGMPmathgrk{s}{\left\LeftPost{par}\SGMPmathgrk{h}\right\RightPost{par}}\)
is
\(
\SGMPmathgrk{s}{\left\LeftPost{par}\SGMPmathgrk{h}\right\RightPost{par}}=ia'/2a\).
The Hamiltonian (23) is precisely of the form of Eq. (14).
In the Schr\"odinger picture, the quantum state (wave-function)
transforms from the vacuum state to a one-mode squeezed vacuum state as
a result of the Schr\"odinger evolution with the Hamiltonian (23). We
will use the Heisenberg picture in which the operators change with time
but the quantum state of the system remains fixed. We also wish to write
the Hamiltonian in the form which would explicitly demonstrate the
underlying physical phenomenon: particle creation in pairs with the
oppositely directed momenta. The Hamiltonian to be used in Eq.~(19) has
the form
\def\XRefId{}
\begin{equation}\SGMPlab\XRefId\vcenter{\halign{\strut\hfil#\hfil&#\hfil\cr
$\displaystyle{H=na^{+}_{{\bf n}}a_{{\bf n}}+n
a^{+}_{-{\bf n}}a_{-{\bf n}}+2\SGMPmathgrk{s}
{\left\LeftPost{par}\SGMPmathgrk{h}\right\RightPost{par}}a^{+}_{%
{\bf n}}a^{+}_{-{\bf n}}+2\SGMPmathgrk{s}^{%
*}{\left\LeftPost{par}\SGMPmathgrk{h}\right\RightPost{par}}a_{{\bf n}
}a_{-{\bf n}}}$\cr
}}\end{equation}

\noindent
This Hamiltonian can be derived from Eq. (23) if one considers the sum
of two Hamiltonians (23) for two modes (with the same frequency
\( n\))
\( b_{1}\) and
\( b_{2}\) and introduces
\( a_{{\bf n}},\hskip 0.167em a_{-{\bf n}}\) according to the relations:
\def\XRefId{}
\begin{equation}\SGMPlab\XRefId\vcenter{\halign{\strut\hfil#\hfil&#\hfil\cr
$\displaystyle{a_{{\bf n}}={{b_{1}-ib_{2}
}\over{\Rad{2}\DoRad }},\hskip 0.265em \hskip 0.265em
\hskip 0.265em a_{-{\bf n}}={{b_{1}+i
b_{2}}\over{\Rad{2}\DoRad }}}$\cr
}}\end{equation}

The descriptions based on the
\( b\)-operators corresponds to standing waves and one-mode
squeezed states while the description based on the
\( a\)-operators corresponds to travelling waves and
two-mode squeezed states.\par

\par By using Eq. (24) one can write the Heisenberg equations of motion
(19) in the form
\def\XRefId{}
\begin{equation}\SGMPlab\XRefId\vcenter{\halign{\strut\hfil#\hfil&#\hfil\cr
$\displaystyle{i{{da_{{\bf n}}}\over{d\SGMPmathgrk{h}}}
=na_{{\bf n}}+i{{a'}\over{a}}a^{+
}_{-{\bf n}},\hskip 0.265em \hskip 0.265em \hskip 0.265em
-i{{da^{+}_{{\bf n}}}\over{d\SGMPmathgrk{h}
}}=na^{+}_{{\bf n}}-i{{a'}
\over{a}}a_{-{\bf n}}}$\cr
}}\end{equation}

\noindent
for each
\( s\) and each \( {\bf n}\).
The solutions to these equations can be written as
\def\XRefId{}
\begin{equation}\SGMPlab\XRefId\vcenter{\halign{\strut\hfil#\hfil&#\hfil\cr
$\displaystyle{}$\hfilneg&$\displaystyle{{}a_{{\bf n}}
{\left\LeftPost{par}\SGMPmathgrk{h}\right\RightPost{par}}
=u_{n}{\left\LeftPost{par}\SGMPmathgrk{h}\right\RightPost{par}}a_{%
{\bf n}}{\left\LeftPost{par}0\right\RightPost{par}}+v_{n}
{\left\LeftPost{par}\SGMPmathgrk{h}\right\RightPost{par}}a^{+}_{%
-{\bf n}}{\left\LeftPost{par}0\right\RightPost{par}}}$\cr
$\displaystyle{a^{+}_{{\bf n}}{\left\LeftPost{par}
\SGMPmathgrk{h}\right\RightPost{par}}
=u^{*}_{n}{\left\LeftPost{par}\SGMPmathgrk{h}\right\RightPost{par}}
a^{+}_{{\bf n}}{\left\LeftPost{par}0\right\RightPost{par}}
+v^{*}_{n}{\left\LeftPost{par}\SGMPmathgrk{h}\right\RightPost{par}}
a_{-{\bf n}}{\left\LeftPost{par}0\right\RightPost{par}}}$\cr
}}\end{equation}

\noindent
where
\(
a_{{\bf n}}{\left\LeftPost{par}0\right\RightPost{par}},\hskip 0.167em
a^{+}_{{\bf n}}{\left\LeftPost{par}0\right\RightPost{par}}
\)
are the initial values of the operators
\(
a_{{\bf n}}{\left\LeftPost{par}\SGMPmathgrk{h}\right\RightPost{par}}
,\hskip 0.167em a^{+}_{{\bf n}}{\left\LeftPost{par}
\SGMPmathgrk{h}\right\RightPost{par}}\)
for some initial time and the complex functions
\(
u_{n}{\left\LeftPost{par}\SGMPmathgrk{h}\right\RightPost{par}},\hskip 0.167em
v_{n}{\left\LeftPost{par}\SGMPmathgrk{h}\right\RightPost{par}}\)
satisfy the equations
\def\XRefId{}
\begin{equation}\SGMPlab\XRefId\vcenter{\halign{\strut\hfil#\hfil&#\hfil\cr
$\displaystyle{i{{du_{n}}\over{d\SGMPmathgrk{h}}}=nu_{%
n}+i{{a'}\over{a}}v^{*}_{n}
,\hskip 0.265em \hskip 0.265em \hskip 0.265em i{{dv_{%
n}}\over{d\SGMPmathgrk{h}}}=nv_{n}+i{{a
'}\over{a}}u^{*}_{n}}$\cr
}}\end{equation}

\noindent
where
\(
{\left\LeftPost{vb}u_{n}\right\RightPost{vb}}^{2}-{\left\LeftPost{vb}
v_{n}\right\RightPost{vb}}^{2}=1\)
and
\(
u_{n}{\left\LeftPost{par}0\right\RightPost{par}}=1,\hskip 0.167em
v_{n}{\left\LeftPost{par}0\right\RightPost{par}}=0\).
Note that
\( u_{n},\hskip 0.167em v_{n}\)
depend only on the absolute value of the vector
\( {\bf n}\),
not its direction. The operators
\(
a_{{\bf n}}{\left\LeftPost{par}0\right\RightPost{par}},\hskip 0.167em
a^{+}_{{\bf n}}{\left\LeftPost{par}0\right\RightPost{par}}
\)
(Schr\"odinger operators) obey the commutation relations
\(
{\left\LeftPost{sqb}a_{{\bf n}}{\left\LeftPost{par}0\right\RightPost{par}}
,\hskip 0.167em a^{+}_{{\bf m}}{\left\LeftPost{par}
0\right\RightPost{par}}\right\RightPost{sqb}}=\SGMPmathgrk{d}^{3}
{\left\LeftPost{par}{\bf n}-{\bf m}\right\RightPost{par}}\)
and so do the evolved operators:
\(
{\left\LeftPost{sqb}a_{{\bf n}}{\left\LeftPost{par}
\SGMPmathgrk{h}\right\RightPost{par}}
,\hskip 0.167em a_{{\bf
m}}{\left\LeftPost{par}\SGMPmathgrk{h}\right\RightPost{par}}
\right\RightPost{sqb}}=\SGMPmathgrk{d}^{3}{\left\LeftPost{par}{\bf n}
-{\bf m}\right\RightPost{par}}\) .\par

\par It follows from Eq. (28) that the function
\( u_{n}+v^{*}_{n}\)
satisfies the equation
\def\XRefId{}
\begin{equation}\SGMPlab\XRefId\vcenter{\halign{\strut\hfil#\hfil&#\hfil\cr
$\displaystyle{{\left\LeftPost{par}u_{n}+v^{*}_{n}\right\RightPost{par}}
''+{\left\LeftPost{par}n^{2}-{{a''}\over{a}}
\right\RightPost{par}}{\left\LeftPost{par}u_{n}+v^{*}_{%
n}\right\RightPost{par}}=0}$\cr
}}\end{equation}

\noindent
which is precisely Eq. (6). The two complex functions
\( u_{n},\hskip 0.167em v_{n}\)
restricted by one constraint
\(
{\left\LeftPost{vb}u_{n}\right\RightPost{vb}}^{2}-{\left\LeftPost{vb}
v_{n}\right\RightPost{vb}}^{2}=1\)
can be parameterized by the three real functions
\(
r_{n}{\left\LeftPost{par}\SGMPmathgrk{h}\right\RightPost{par}},\hskip 0.167em
\SGMPmathgrk{f}_{n}{\left\LeftPost{par}
\SGMPmathgrk{h}\right\RightPost{par}},\hskip 0.167em
\SGMPmathgrk{q}_{n}{\left\LeftPost{par}\SGMPmathgrk{h}\right\RightPost{par}}
\):
\def\XRefId{}
\begin{equation}\SGMPlab\XRefId\vcenter{\halign{\strut\hfil#\hfil&#\hfil\cr
$\displaystyle{u=e^{i\SGMPmathgrk{q}}chr,\hskip 0.265em \hskip 0.265em
\hskip 0.265em v=e^{-i{\left\LeftPost{par}
\SGMPmathgrk{q}-2\SGMPmathgrk{f}\right\RightPost{par}}
}shr}$\cr
}}\end{equation}

\noindent
which are, corrrespondingly, squeeze parameter
\( r\), squeeze angle
\( \SGMPmathgrk{f}\) and rotation angle
\( \SGMPmathgrk{q}\). For each
\( n\) and \( s\), they obey the equations
\def\XRefId{}
\begin{equation}\SGMPlab\XRefId\vcenter{\halign{\strut\hfil#\hfil&#\hfil\cr
$\displaystyle{r'={{a'}\over{a}}\cos 2\SGMPmathgrk{f}}$\cr
$\displaystyle{\SGMPmathgrk{q}'=n-{{a'}\over{a}}\sin 2\SGMPmathgrk{f}
\hskip 0.265em thr}$\cr
$\displaystyle{\SGMPmathgrk{f}'=-n-{{a'}\over{a}}\sin 2\SGMPmathgrk{f}
\hskip 0.265em cthr}$\cr
}}\end{equation}

\noindent
which can be used for an explicit calculation of
\(
r,\hskip 0.167em \SGMPmathgrk{f},\hskip 0.167em \SGMPmathgrk{q}\)
for a given scale factor
\(
a{\left\LeftPost{par}\SGMPmathgrk{h}\right\RightPost{par}}\).\par

\par The squeeze parameters have been calculated [17] (by a different
method) for a model which we have considered in Sec. 5. The model
includes three successive stages of expansion: De-Sitter,
radiation-dominated and matter-dominated. It has been shown that the
squeeze parameter
\( r\) varies in the large interval from
\( r{\ifmmode\approx\else$\approx$\fi}1\) to
\( r{\ifmmode\approx\else$\approx$\fi}120\)
over the spectrum of relic gravtational waves (see
Fig. (4)). The value
\(
r{\ifmmode\approx\else$\approx$\fi}1\)
applies to the shortest waves with the present day
frequencies of order of
\(
\SGMPmathgrk{n}{\ifmmode\approx\else$\approx$\fi}10^{8}\) Hz,
\(
r{\ifmmode\approx\else$\approx$\fi}10^{2}\)
is the value of
\( r\) attributed to the waves with frequencies
\(
\SGMPmathgrk{n}{\ifmmode\approx\else$\approx$\fi}10^{-16}\) Hz, and
\( r{\ifmmode\approx\else$\approx$\fi}120\)
corresponds to the waves with the Hubble frequencies
\(
\SGMPmathgrk{n}{\ifmmode\approx\else$\approx$\fi}10^{-18}\) Hz.\par

\par It is important to note that Eqs. (27) can be cast in the form
\def\XRefId{}
\begin{equation}\SGMPlab\XRefId\vcenter{\halign{\strut\hfil#\hfil&#\hfil\cr
$\displaystyle{a_{{\bf n}}
{\left\LeftPost{par}\SGMPmathgrk{h}\right\RightPost{par}}
=RSa_{{\bf n}}{\left\LeftPost{par}0\right\RightPost{par}}S^{%
+}R^{+},\hskip 0.265em \hskip 0.265em \hskip 0.265em
a^{+}_{{\bf n}}{\left\LeftPost{par}\SGMPmathgrk{h}\right\RightPost{par}}
=RSa^{+}_{{\bf n}}{\left\LeftPost{par}0\right\RightPost{par}}
S^{+}R^{+}}$\cr
}}\end{equation}
where
\def\XRefId{}
\begin{equation}\SGMPlab\XRefId\vcenter{\halign{\strut\hfil#\hfil&#\hfil\cr
$\displaystyle{S{\left\LeftPost{par}r,
\hskip 0.167em \SGMPmathgrk{f}\right\RightPost{par}}
={\rm exp}{\left\LeftPost{sqb}r{\left\LeftPost{par}e^{-2i\SGMPmathgrk{f}}
a_{{\bf n}}{\left\LeftPost{par}0\right\RightPost{par}}a_{%
-{\bf n}}{\left\LeftPost{par}0\right\RightPost{par}}-e^{2i\SGMPmathgrk{f}
}a^{+}_{{\bf n}}{\left\LeftPost{par}0\right\RightPost{par}}
a^{+}_{-{\bf n}}{\left\LeftPost{par}0\right\RightPost{par}}
\right\RightPost{par}}\right\RightPost{sqb}}}$\cr
}}\end{equation}

\noindent
is the unitary two-mode squeeze operator and
\def\XRefId{}
\begin{equation}\SGMPlab\XRefId\vcenter{\halign{\strut\hfil#\hfil&#\hfil\cr
$\displaystyle{R{\left\LeftPost{par}
\SGMPmathgrk{q}\right\RightPost{par}}={\rm exp}
{\left\LeftPost{sqb}-i\SGMPmathgrk{q}{\left\LeftPost{par}a^{+}_{{\bf n}
}{\left\LeftPost{par}0\right\RightPost{par}}a_{{\bf n}}
{\left\LeftPost{par}0\right\RightPost{par}}+a^{+}_{-{\bf n}
}{\left\LeftPost{par}0\right\RightPost{par}}a_{-{\bf n}}
{\left\LeftPost{par}0\right\RightPost{par}}
\right\RightPost{par}}\right\RightPost{sqb}}
}$\cr
}}\end{equation}

\noindent
is the unitary rotation operator. Eq. (32) demonstartes explicitly the
inevitable appearance of squeezing in this kind of problem.\par

\par We will assume that the quantum state of the field is the vacuum
state defined by the requirement
\(
a_{{\bf n}}{\left\LeftPost{par}0\right\RightPost{par}}{\ifmmode |\else$ |$\fi}
0{\ifmmode>\else$>$\fi}=0\)
for each
\( {\bf n}\) and for both
\( s\). The values of
\(
a_{{\bf n}}{\left\LeftPost{par}\SGMPmathgrk{h}\right\RightPost{par}}
,\hskip 0.167em a^{+}_{{\bf n}}{\left\LeftPost{par}
\SGMPmathgrk{h}\right\RightPost{par}}\)
determine all the statistical properties of the field
in the later times. The mean values of
\(
a_{{\bf n}}{\left\LeftPost{par}\SGMPmathgrk{h}\right\RightPost{par}}
,\hskip 0.167em a^{+}_{{\bf n}}{\left\LeftPost{par}
\SGMPmathgrk{h}\right\RightPost{par}}\)
are zero:
\(
{\ifmmode<\else$<$\fi}0{\ifmmode |\else$ |$\fi}
a_{{\bf n}}{\left\LeftPost{par}\SGMPmathgrk{h}\right\RightPost{par}}
{\ifmmode |\else$ |$\fi}0{\ifmmode>\else$>$\fi}=0,\hskip 0.265em \hskip 0.265em
\hskip 0.265em
{\ifmmode<\else$<$\fi}0
{\ifmmode |\else$ |$\fi}a^{+}_{{\bf n}}{\left\LeftPost{par}\SGMPmathgrk{h}
\right\RightPost{par}}{\ifmmode |\else$ |$\fi}0{\ifmmode>\else$>$\fi}=0\).
The mean values of the quadratic combinations of
\(
a_{{\bf n}}{\left\LeftPost{par}\SGMPmathgrk{h}\right\RightPost{par}}
,\hskip 0.167em a^{+}_{{\bf n}}{\left\LeftPost{par}
\SGMPmathgrk{h}\right\RightPost{par}}\) are not zero:
\def\XRefId{}
\begin{equation}\SGMPlab\XRefId\vcenter{\halign{\strut\hfil#\hfil&#\hfil\cr
$\displaystyle{}$\hfilneg&$\displaystyle{{}
{\ifmmode<\else$<$\fi}0{\ifmmode |\else$ |$\fi}a_{{\bf n}}{\left\LeftPost{par}
\SGMPmathgrk{h}\right\RightPost{par}}a_{{\bf m}}{\left\LeftPost{par}
\SGMPmathgrk{h}\right\RightPost{par}}{\ifmmode |\else$
|$\fi}0{\ifmmode>\else$>$\fi}=u_{n}{\left\LeftPost{par}
\SGMPmathgrk{h}\right\RightPost{par}}v_{m}{\left\LeftPost{par}\SGMPmathgrk{h}
\right\RightPost{par}}\SGMPmathgrk{d}^{3}{\left\LeftPost{par}{\bf n}
+{\bf m}\right\RightPost{par}}}$\cr
$\displaystyle{}$\hfilneg&$\displaystyle{{}{\ifmmode<\else$<$\fi}0
{\ifmmode |\else$ |$\fi}a^{+}_{{\bf n}}
{\left\LeftPost{par}\SGMPmathgrk{h}\right\RightPost{par}}a^{+}_{%
{\bf m}}{\left\LeftPost{par}
\SGMPmathgrk{h}\right\RightPost{par}}{\ifmmode |\else$ |$\fi}
0{\ifmmode>\else$>$\fi}=v^{*}_{n}{\left\LeftPost{par}
\SGMPmathgrk{h}\right\RightPost{par}}
u^{*}_{m}{\left\LeftPost{par}
\SGMPmathgrk{h}\right\RightPost{par}}
\SGMPmathgrk{d}^{3}{\left\LeftPost{par}{\bf n}+
{\bf m}\right\RightPost{par}}
}$\cr
$\displaystyle{}$\hfilneg&
$\displaystyle{{}{\ifmmode<\else$<$\fi}0{\ifmmode |\else$ |$\fi}
a_{{\bf n}}{\left\LeftPost{par}
\SGMPmathgrk{h}\right\RightPost{par}}a^{+}_{{\bf m}}
{\left\LeftPost{par}\SGMPmathgrk{h}\right\RightPost{par}}
{\ifmmode |\else$ |$\fi}0{\ifmmode>\else$>$\fi}=u_{n
}{\left\LeftPost{par}\SGMPmathgrk{h}\right\RightPost{par}}u^{*}_{%
m}{\left\LeftPost{par}\SGMPmathgrk{h}\right\RightPost{par}}\SGMPmathgrk{d}^{%
3}{\left\LeftPost{par}{\bf n}-{\bf m}\right\RightPost{par}}}$\cr
$\displaystyle{}$\hfilneg&$\displaystyle{{}{\ifmmode<\else$<$\fi}0
{\ifmmode |\else$ |$\fi}a^{+}_{{\bf n}}
{\left\LeftPost{par}\SGMPmathgrk{h}\right\RightPost{par}}a_{{\bf m}}
{\left\LeftPost{par}\SGMPmathgrk{h}\right\RightPost{par}}
{\ifmmode |\else$ |$\fi}0{\ifmmode>\else$>$\fi}=v^{*
}_{{\bf n}}{\left\LeftPost{par}\SGMPmathgrk{h}\right\RightPost{par}}
v_{m}{\left\LeftPost{par}\SGMPmathgrk{h}\right\RightPost{par}}
\SGMPmathgrk{d}^{%
3}{\left\LeftPost{par}{\bf n}-{\bf m}\right\RightPost{par}}}$\cr
}}\end{equation}

\par

\par These relationships (first two) show explicitly that the waves
(modes) with the opposite momenta are not independent but, on the
contrary, are strongly correlated. This means that the generated field
is a combination of standing waves [17]. Let us see how this is
reflected in the correlation functions of the field.\par

\par For purposes of illustration, we will first ignore the tensorial
indices in Eq.~(18) and will consider a scalar field
\def\XRefId{}
\begin{equation}\SGMPlab\XRefId\vcenter{\halign{\strut\hfil#\hfil&#\hfil\cr
$\displaystyle{h{\left\LeftPost{par}\SGMPmathgrk{h},\hskip 0.167em
{\bf x}\right\RightPost{par}}
=\int _{-\infty }^{\infty }d^{3}
{\bf n}{\left\LeftPost{par}a_{{\bf n}}{\left\LeftPost{par}\SGMPmathgrk{h}
\right\RightPost{par}}e^{i{\bf nx}}+a^{+}_{%
{\bf n}}{\left\LeftPost{par}\SGMPmathgrk{h}\right\RightPost{par}}e^{%
-i{\bf nx}}\right\RightPost{par}}}$\cr
}}\end{equation}

\noindent
It is obvious that the mean value of the field is
zero,
\(
{\ifmmode<\else$<$\fi}0{\ifmmode |\else$
|$\fi}h{\left\LeftPost{par}\SGMPmathgrk{h},\hskip 0.167em {\bf
x}\right\RightPost{par}}
{\ifmmode |\else$ |$\fi}0{\ifmmode>\else$>$\fi}=0\),
in every spatial point and for every moment of time.
The mean value of the square of the field
\(
h{\left\LeftPost{par}{\bf \SGMPmathgrk{h}},\hskip 0.167em {\bf
x}\right\RightPost{par}}
\)
is not zero and can be calculated with the help of
Eq.~(35):\par

\def\XRefId{}

\par
\def\XRefId{}
\begin{equation}\SGMPlab\XRefId\vcenter{\halign{\strut\hfil#\hfil&#\hfil\cr
$\displaystyle{{\ifmmode<\else$<$\fi}0{\ifmmode |\else$ |$\fi}
h{\left\LeftPost{par}\SGMPmathgrk{h},\hskip 0.167em {\bf x}
\right\RightPost{par}},\hskip 0.167em h{\left\LeftPost{par}\SGMPmathgrk{h}
,\hskip 0.167em {\bf x}\right\RightPost{par}}{\ifmmode |\else$ |$\fi}
0{\ifmmode>\else$>$\fi}=4\SGMPmathgrk{p}
\int _{0}^{\infty }n^{2}dn{\left\LeftPost{par}
{\left\LeftPost{vb}u_{n}\right\RightPost{vb}}^{2}+{\left\LeftPost{vb}
v_{n}\right\RightPost{vb}}^{2}+u_{n}v_{%
n}+u^{*}_{n}v^{*}_{n}\right\RightPost{par}}
}$\cr
}}\end{equation}

\noindent
In term of the squeeze parameters this expression can be written as
\def\XRefId{}
\begin{equation}\SGMPlab\XRefId\vcenter{\halign{\strut\hfil#\hfil&#\hfil\cr
$\displaystyle{{\ifmmode<\else$<$\fi}0
{\ifmmode |\else$ |$\fi}h{\left\LeftPost{par}\SGMPmathgrk{h},\hskip 0.167em
{\bf x}
\right\RightPost{par}},\hskip 0.167em h{\left\LeftPost{par}\SGMPmathgrk{h}
,\hskip 0.167em {\bf x}\right\RightPost{par}}
{\ifmmode |\else$ |$\fi}0{\ifmmode>\else$>$\fi}=4\SGMPmathgrk{p}
\int _{0}^{\infty }n^{2}dn{\left\LeftPost{par}
ch2r_{n}+sh2r_{n}\cdot{\cos 2\SGMPmathgrk{f}_{%
n}}\right\RightPost{par}}}$\cr
}}\end{equation}

\noindent
(it includes the vacuum term
\( 4\SGMPmathgrk{p}\int _{0}^{\infty }n^{2 }dn\),
which should be subtracted at the end). It is seen
from Eq.~(38) that the variance of the field does not depend on the
spatial coordinate
\( {\bf x}\).
The function under the integral in Eq.~(38) is
usually called the power spectrum of the field:
\(
P{\left\LeftPost{par}n\right\RightPost{par}}=n^{2}{\left\LeftPost{par}
ch2r_{n}+sh2r_{n}\cos 2\SGMPmathgrk{f}_{n}
\right\RightPost{par}}\).
The important property of squeezing is that the
\(
P{\left\LeftPost{par}n\right\RightPost{par}}\)
is not a smooth function of \( n\)
but is modulated and contains many zeros or, strictly
speaking, very deep minima. To see this, one can return to Eq.~(31). For
late times, that is, well after the completion of the amplification
process, the function
\( a'/a\)
on the right-hand side of Eq.~(31) can be neglected.
This is equivalent to saying that one is considering waves that are well
inside the present day Hubble radius. For these late times, the squeeze
parameter
\( r_{n}\) is not growing any more and the squeeze angle
\( \SGMPmathgrk{f}_{n}\) is just
\(
\SGMPmathgrk{f}_{n}=-n\SGMPmathgrk{h}-\SGMPmathgrk{f}_{0n}\).
Since
\(
r_{n}{\ifmmode\gg\else$\gg$\fi}1\)
for the frequencies of our interest, the
\(
P{\left\LeftPost{par}n\right\RightPost{par}}\)
can be written as
\(
P{\left\LeftPost{par}n\right\RightPost{par}}
{\ifmmode\approx\else$\approx$\fi}n^{2}e^{2
r_{n}}\SGMPlim{r}\cos \SGMPdolim ^{2}
{\left\LeftPost{par}n\SGMPmathgrk{h}+\SGMPmathgrk{f}_{0n}\right\RightPost{par}}
\).
The factor
\(
\SGMPlim{r}\cos \SGMPdolim ^{2}{\left\LeftPost{par}
n\SGMPmathgrk{h}+\SGMPmathgrk{f}_{0n}\right\RightPost{par}}\)
vanishes for a series of values of
\( n\). At these frequencies, the function
\( P{\left\LeftPost{par}n\right\RightPost{par}}\)
goes to zero. The position of zeros on the
\( n\) axis varies with time. Similar conclusions hold for
the spatial correlation function
\(
{\ifmmode<\else$<$\fi}0{\ifmmode |\else$ |$\fi}n
{\left\LeftPost{par}\SGMPmathgrk{h},\hskip 0.167em {\bf
x}\right\RightPost{par}}
,\hskip 0.167em h{\left\LeftPost{par}\SGMPmathgrk{h},\hskip 0.167em {\bf x
}+{\bf l}\right\RightPost{par}}{\ifmmode |\else$
|$\fi}0{\ifmmode>\else$>$\fi}\)
:
\def\XRefId{}
\begin{equation}\SGMPlab\XRefId\vcenter{\halign{\strut\hfil#\hfil&#\hfil\cr
$\displaystyle{{\ifmmode<\else$<$\fi}0{\ifmmode |\else$ |$\fi}h
{\left\LeftPost{par}\SGMPmathgrk{h},\hskip 0.167em {\bf x}
\right\RightPost{par}},\hskip 0.167em h{\left\LeftPost{par}\SGMPmathgrk{h}
,\hskip 0.167em {\bf x}+{\bf l}\right\RightPost{par}}{\ifmmode |\else$ |$\fi}0
{\ifmmode>\else$>$\fi}=4\SGMPmathgrk{p}\int _{0}^{\infty }n^{%
2}{{\sin nl}\over{nl}}{\left\LeftPost{par}
ch2r_{n}+sh2r_{n}\cos 2\SGMPmathgrk{f}_{n}
\right\RightPost{par}}dn}$\cr
}}\end{equation}

\noindent
The resulting expression (39) depends on the distance between the
spatial points but not on their coordinates. The power spectrum of this
correlation function is also modulated by the same factor
\(
\SGMPlim{r}\cos \SGMPdolim ^{2}{\left\LeftPost{par}
n\SGMPmathgrk{h}+\SGMPmathgrk{f}_{0n}\right\RightPost{par}}\).
It is necessary to note that the power spectrum of
the energy density of the field is smoooth as it includes, in addition
to Eq.~(39), the kinetic energy term
\def\XRefId{}
\begin{equation}\SGMPlab\XRefId\vcenter{\halign{\strut\hfil#\hfil&#\hfil\cr
$\displaystyle{{{1}\over{n^{2}}}{\ifmmode<\else$<$\fi}0
{\ifmmode |\else$ |$\fi}h'{\left\LeftPost{par}
\SGMPmathgrk{h},\hskip 0.167em {\bf
x}\right\RightPost{par}}h'{\left\LeftPost{par}
\SGMPmathgrk{h},\hskip 0.167em {\bf x}\right\RightPost{par}}
{\ifmmode |\else$ |$\fi}0{\ifmmode>\else$>$\fi}
=4\SGMPmathgrk{p}\int _{0}^{\infty }n^{%
2}dn{\left\LeftPost{par}ch2r_{n}-sh2r_{n}\cos
2\SGMPmathgrk{f}_{n}\right\RightPost{par}}}$\cr
}}\end{equation}

\noindent
so the contributions with the oscillating factors
\( \cos 2\SGMPmathgrk{f}_{n}\) cancel out.\par

\par Let us mention some new features which arise when one considers the
tensor field
\(
h_{ij}{\left\LeftPost{par}\SGMPmathgrk{h},\hskip 0.167em {\bf x}
\right\RightPost{par}}\),
Eq.~(18), not the scalar field
\(
h{\left\LeftPost{par}\SGMPmathgrk{h},\hskip 0.167em {\bf
x}\right\RightPost{par}}
\).
First, one sees from Eq.~(28) that the solutions
\(
u^{s}_{n}{\left\LeftPost{par}\SGMPmathgrk{h}\right\RightPost{par}}
,\hskip 0.167em v^{s}_{n}{\left\LeftPost{par}\SGMPmathgrk{h}
\right\RightPost{par}}\)
for both polarisations
\( s=1,\hskip 0.167em 2\)
are identical; they obey the same equations with the
same initial conditions. This means that, for each
\( {\bf n}\), both polarisation states are necessarily generated,
and with equal amplitudes. This feature can serve as a clear distinction
between gravitational waves generated quantum-mechanically and by other
mechanisms. The second feature is related to the properties of the
correlation function analogous to Eq.~(39). In case of the tensor field
\( h_{ij}\), there is one combination of the components
\( h_{ij}\) which has a particular interest:
\( h{\left\LeftPost{par}e\right\RightPost{par}}=h_{ij}e^{i} e^{j}\),
where
\( e^{i}\) is an arbitrary unit vector. The
\( h{\left\LeftPost{par}e\right\RightPost{par}}\)
enters the calculations of the fluctuations of the
microwave background temperature seen in the direction
\( e^{i}\) (Sachs-Wolfe effect):
\def\XRefId{}
\begin{equation}\SGMPlab\XRefId\vcenter{\halign{\strut\hfil#\hfil&#\hfil\cr
$\displaystyle{{{\bigtriangleup T}\over{T}}{\left\LeftPost{par}e^{%
i}\right\RightPost{par}}={{1}\over{2}}\int
_{\SGMPmathgrk{h}_{E}}^{\SGMPmathgrk{h}_{R}}
{{\SGMPmathgrk{6}h_{ij}}\over{\SGMPmathgrk{6}\SGMPmathgrk{h}}}
e^{i}e^{j}d\SGMPmathgrk{h}}$\cr
}}\end{equation}

The relevant correlation function is
\(
{\ifmmode<\else$<$\fi}0{\ifmmode |\else$
|$\fi}h_{ij}e^{i}e^{j}{\left\LeftPost{par}
\SGMPmathgrk{h},\hskip 0.167em 0\right\RightPost{par}}h_{ij}
e^{i}e^{j}{\left\LeftPost{par}\SGMPmathgrk{h},\hskip 0.167em
\SGMPmathgrk{t}e^{k}\right\RightPost{par}}
{\ifmmode |\else$ |$\fi}0{\ifmmode>\else$>$\fi}\),
where
\( \SGMPmathgrk{t}\) is a parameter along the line of sight.
It is interesting to calculate this function for the initial time
\( \SGMPmathgrk{h}=0\) and, also, for some very late time
\( \SGMPmathgrk{h}\). Without going into the details (they will be
published elsewhere), we will present some results. For
\( \SGMPmathgrk{h}=0\) one has
\(
v_{n}{\left\LeftPost{par}\SGMPmathgrk{h}\right\RightPost{par}}=0,\hskip 0.167em
u_{n}{\left\LeftPost{par}\SGMPmathgrk{h}\right\RightPost{par}}=1\)
and
\def\XRefId{}
\begin{equation}\SGMPlab\XRefId\vcenter{\halign{\strut\hfil#\hfil&#\hfil\cr
$\displaystyle{{\ifmmode<\else$<$\fi}0
{\ifmmode |\else$ |$\fi}h_{ij}e^{i}e^{j}{\left\LeftPost{par}
0,\hskip 0.167em 0\right\RightPost{par}}h_{ij}e^{i
}e^{j}{\left\LeftPost{par}0,\hskip 0.167em \SGMPmathgrk{t}e^{%
k}\right\RightPost{par}}{\ifmmode |\else$ |$\fi}0
{\ifmmode>\else$>$\fi}=\int _{-\infty
}^{\infty }d^{3}{\bf n}{\left\LeftPost{sqb}
{\left\LeftPost{par}p^{1}_{ij}e^{i}e^{j}
\right\RightPost{par}}^{2}+{\left\LeftPost{par}p^{2}_{%
ij}e^{i}e^{j}\right\RightPost{par}}^{2
}\right\RightPost{sqb}}e^{-in_{j}e^{j}\tau }}$\cr
}}\end{equation}

The presence of the both polarisations is very important: because of
this the integration over angular variables eliminates dependence on the
direction
\( e^{j}\). The final result is
\def\XRefId{}
\begin{equation}\SGMPlab\XRefId\vcenter{\halign{\strut\hfil#\hfil&#\hfil\cr
$\displaystyle{{\ifmmode<\else$<$\fi}0{\ifmmode |\else$ |$\fi}
h_{ij}e^{i}e^{j}{\left\LeftPost{par}
0,\hskip 0.167em 0\right\RightPost{par}}h_{ij}e^{i
}e^{j}{\left\LeftPost{par}0,\hskip 0.167em \SGMPmathgrk{t}e^{%
k}\right\RightPost{par}}{\ifmmode |\else$ |$\fi}0{\ifmmode>\else$>$\fi}=
16\SGMPmathgrk{p}\Rad{{{%
\SGMPmathgrk{p}}\over{2}}}\DoRad \int _{0}
^{\infty }{\left\LeftPost{par}n\SGMPmathgrk{t}\right\RightPost{par}}^{%
-5/2}J_{5/2}{\left\LeftPost{par}n\SGMPmathgrk{t}\right\RightPost{par}}
n^{2}dn}$\cr
}}\end{equation}

\noindent
One can see that the (vacuum) correlation function depends only on the
distance between the points and does not depend on the direction
\( e^{j}\).\par

\par To conclude this section, we should say that the same theory of
squeezed states is applicable to the density perturbations generated
quantum-mechanically [18]. In a simple inflationary model governed by a
massive scalar field, the progenitor of the density perturbations and,
later, the density perturbations themselves, satisfy equations similar
to the equations for gravitational waves:
\(
v''+{\left\LeftPost{par}n^{2}-V{\left\LeftPost{par}n,\hskip 0.167em
\SGMPmathgrk{h}\right\RightPost{par}}\right\RightPost{par}}v=0\)
where
\( v\) is a gauge-invariant function which includes
perturbations of the matter variables and gravitational field [19], and
\(
V{\left\LeftPost{par}n,\hskip 0.167em \SGMPmathgrk{h}\right\RightPost{par}}
\)
is the effective potential analogous to the gravitational-wave potential
\(
U{\left\LeftPost{par}\SGMPmathgrk{h}\right\RightPost{par}}\),
Fig.~3. Similarly to the case of gravitational
waves, squeezing in density perturbations and associated (longitudinal)
gravitational field, exhibits distinct observational
features.\SGMPnewline
\par

\par \underline{{\protect\normalsize {\bf 8.
Quantum Cosmology, Minisuperspace Models and
Inflation}}}\par

\par Until now we have been discussing the quantum fluctuations
superimposed on a given classical background spacetime. The next level
of complexity is the quantization of the background geometry itself.
This is the domain of full quantum gravity and quantum cosmology.\par

\par In canonical quantum gravity the role of a generalized coordinate
is played by a 3{--}geometry
\( g^{{\left\LeftPost{par}3\right\RightPost{par}}}\).
The full set of all 3{--}geometries forms a
superspace, where the wave function of the quantized gravitational field
is defined. If some matter fields are present, the superspace includes
the matter variables as well. In cosmological applications, one usually
considers topologically compact geometries and calls the wave function
of the entire system the wave function of the universe. The basic
equation which governs the wave function of the universe is called the
Wheeler {--} DeWitt (WD) equation. (For reviews of quantum gravity
and quantum cosmology, see, for example [20].)\par

\par A simplified problem, which allows a detailed investigation, is
provided by minisuperspace models. In minisuperspace models one neglects
all degrees of freedom except of a few. A quantum cosmological model
describing a homogeneous isotropic universe filled with a massive scalar
field gives a reasonably simple, though sufficiently representative
case. In this case one has only two degrees of freedom (two
minisuperspace variables): the scale factor a(t) and the scalar field
$\phi$(t). Since the formulation of a quantum problem includes
integration of some quantities, such as the Hamiltonian function, over 3
{--} volume, one normally considers closed 3{--}sphere geometries,
k = +1, or torus-like geometries, k=0, in order to avoid infinities
arising because of spatial integration. The total energy of a closed
world is zero. This is why the analog of the Schr\"odinger equation
takes the form \^H$\psi$ = 0, which is the Wheeler {--} DeWitt
equation.\par

\par For a FRW universe filled with a scalar field $\phi$,
\(
V{\left\LeftPost{par}\SGMPmathgrk{f}\right\RightPost{par}}={{1}\over{%
2}}m^{2}\SGMPmathgrk{f}^{2}\),
the Wheeler {--} DeWitt equation can be written
as follows [21]:
\def\XRefId{}
\begin{equation}\SGMPlab\XRefId\vcenter{\halign{\strut\hfil#\hfil&#\hfil\cr
$\displaystyle{{\left\LeftPost{par}{{1}\over{a^{p}}}{{%
\SGMPmathgrk{6}}\over{\SGMPmathgrk{6}a}}a^{p}{{\SGMPmathgrk{6}
}\over{\SGMPmathgrk{6}a}}-{{1}\over{a^{2}}}
{{\SGMPmathgrk{6}^{2}}\over{\SGMPmathgrk{6}\SGMPmathgrk{f}^{2}
}}-ka^{2}+m^{2}\SGMPmathgrk{f}^{2}a^{%
4}\right\RightPost{par}}\SGMPmathgrk{y}
{\left\LeftPost{par}a,\SGMPmathgrk{f}\right\RightPost{par}}
=0}$\cr
}}\end{equation}

The factor p reflects some ambiguity in the choice of operator ordering.
The possible preferred choice of p for the given model is p = 1.\par

\par First, we will show how classical Einstein equations of motion
follow from the quantum equation (44) in the quasi-classical
approximation. For simplicity we consider the limit where the spatial
curvature term ka${}^{2}$ can be neglected. In this limit (and
for $p = 1$) Eq.~(44) reduces to
\def\XRefId{}
\begin{equation}\SGMPlab\XRefId\vcenter{\halign{\strut\hfil#\hfil&#\hfil\cr
$\displaystyle{{\left\LeftPost{par}{{1}\over{a}}{{\SGMPmathgrk{6}
}\over{\SGMPmathgrk{6}a}}a{{\SGMPmathgrk{6}}\over{\SGMPmathgrk{6}
a}}-{{1}\over{a^{2}}}{{%
\SGMPmathgrk{6}^{2}}\over{\SGMPmathgrk{6}\SGMPmathgrk{f}^{2}}}
+m^{2}\SGMPmathgrk{f}^{2}a^{4}\right\RightPost{par}}
\SGMPmathgrk{y}{\left\LeftPost{par}a,\SGMPmathgrk{f}\right\RightPost{par}}=0}$\cr
}}\end{equation}

\par

\par In the quasi-classical approximation, the wave function
$\psi$(
\( a\), $\phi$) has the form
\(
\SGMPmathgrk{y}{\left\LeftPost{par}a,
\SGMPmathgrk{f}\right\RightPost{par}}=exp{\left\LeftPost{par}
iS{\left\LeftPost{par}a,\SGMPmathgrk{f}\right\RightPost{par}}+
i\SGMPmathgrk{s}{\left\LeftPost{par}
a,\SGMPmathgrk{f}\right\RightPost{par}}+...\right\RightPost{par}}\).
By using this representation, the following
equations can be derived from Eq (45):
\def\XRefId{}
\begin{equation}\SGMPlab\XRefId\vcenter{\halign{\strut\hfil#\hfil&#\hfil\cr
$\displaystyle{-{\left\LeftPost{par}{{\SGMPmathgrk{6}S}\over{\SGMPmathgrk{6}a}}
\right\RightPost{par}}^{2}+{{1}\over{a^{2}
}}{\left\LeftPost{par}{{\SGMPmathgrk{6}S}\over{\SGMPmathgrk{6}\SGMPmathgrk{f}
}}\right\RightPost{par}}^{2}+m^{2}\SGMPmathgrk{f}^{%
2}a^{4}=0}$\cr
}}\end{equation}

\def\XRefId{}
\begin{equation}\SGMPlab\XRefId\vcenter{\halign{\strut\hfil#\hfil&#\hfil\cr
$\displaystyle{i{{\SGMPmathgrk{6}^{2}S}\over{\SGMPmathgrk{6}a^{2}
}}-2{{\SGMPmathgrk{6}S}\over{\SGMPmathgrk{6}a}}\hskip 0.167em
{{\SGMPmathgrk{6}\SGMPmathgrk{s}}\over{\SGMPmathgrk{6}a}}\hskip 0.167em
+{{i}\over{a}}\hskip 0.167em {{\SGMPmathgrk{6}
S}\over{\SGMPmathgrk{6}a}}-{{i}\over{a^{2}
}}\hskip 0.167em {{\SGMPmathgrk{6}^{2}S}\over{%
\SGMPmathgrk{6}\SGMPmathgrk{f}^{2}}}+{{2}\over{a^{%
2}}}\hskip 0.167em {{\SGMPmathgrk{6}S}\over{\SGMPmathgrk{6}
\SGMPmathgrk{f}}}\hskip 0.167em {{\SGMPmathgrk{6}\SGMPmathgrk{s}}
\over{\SGMPmathgrk{6}\SGMPmathgrk{f}}}=0}$\cr
}}\end{equation}

Eq. (46) is the Hamiltonian-Jacobi equation for the action S. A real
solution to Eq. (46), which describes the classical dynamics of the
model, can be presented in the form
\def\XRefId{}
\begin{equation}\SGMPlab\XRefId\vcenter{\halign{\strut\hfil#\hfil&#\hfil\cr
$\displaystyle{S{\left\LeftPost{par}a,\SGMPmathgrk{f}\right\RightPost{par}}=-a^{3}
f{\left\LeftPost{par}\SGMPmathgrk{f}\right\RightPost{par}}}$\cr
}}\end{equation}

\noindent
where the function
\(
f{\left\LeftPost{par}\SGMPmathgrk{f}\right\RightPost{par}}\)
, as a consequence of Eq. (46), satisfies the ordinary
differential equation
\def\XRefId{}
\begin{equation}\SGMPlab\XRefId\vcenter{\halign{\strut\hfil#\hfil&#\hfil\cr
$\displaystyle{9f^{2}-{\left\LeftPost{par}{{df}\over{d\SGMPmathgrk{f}
}}\right\RightPost{par}}^{2}=m^{2}\SGMPmathgrk{f}^{%
2}}$\cr
}}\end{equation}

\par

\par The classical equations of motion can be obtained from Eq. (48) and
the Lagrangian
\(
L={{1}\over{2}}{\left\LeftPost{par}-a\MthAcnt {a}{\dot }^{%
2}+a^{3}\MthAcnt {\SGMPmathgrk{f}}{\dot }-a^{%
3}m^{2}\SGMPmathgrk{f}^{2}\right\RightPost{par}}
\)
of the system in the usual way. One writes
\(
{{\SGMPmathgrk{6}L}\over{\SGMPmathgrk{6}\MthAcnt {a}{\dot }
}}=-a\MthAcnt {a}{\dot }={{\SGMPmathgrk{6}S
}\over{\SGMPmathgrk{6}a}},\hskip 0.265em {{\SGMPmathgrk{6}
L}\over{\SGMPmathgrk{6}\MthAcnt {\SGMPmathgrk{f}}{\dot }}}
=a^{3}\MthAcnt {\SGMPmathgrk{f}}{\dot }={{\SGMPmathgrk{6}
S}\over{\SGMPmathgrk{6}\SGMPmathgrk{f}}}\)
which leads to the relations
\def\XRefId{}
\begin{equation}\SGMPlab\XRefId\vcenter{\halign{\strut\hfil#\hfil&#\hfil\cr
$\displaystyle{\MthAcnt {{{a}\over{a}}}{\dot }
=3f,\hskip 0.265em \hskip 0.265em \hskip 0.265em \hskip 0.265em
\hskip 0.265em \hskip 0.265em \hskip 0.265em \MthAcnt {\SGMPmathgrk{f}
}{\dot }=-f^{{\ifmmode{}^\prime\else${}^\prime$\fi}}}$\cr
}}\end{equation}

\noindent
where the prime denotes a derivative with respect to
\( \SGMPmathgrk{f}\),
and the dot denotes a derivative with respect to the
time
\( t\). By taking time derivatives of Eq. (50) and using Eq.~(49),
one derives the equations of motion which can be cast in the usual
classical form:
\def\XRefId{}
\begin{equation}\SGMPlab\XRefId\vcenter{\halign{\strut\hfil#\hfil&#\hfil\cr
$\displaystyle{\vcenter{\halign{\strut\hfil#\hfil&#\hfil\cr
 $\displaystyle{\MthAcnt {\SGMPmathgrk{f}}{\ddot }+3{{%
\MthAcnt {a}{\dot }}\over{a}}\MthAcnt {\SGMPmathgrk{f}
}{\dot }+m^{2}\SGMPmathgrk{f}=0,\hskip 0.167em
{\left\LeftPost{par}{{\MthAcnt {a}{\dot }}\over{%
a}}\right\RightPost{par}}^{2}=\MthAcnt {\SGMPmathgrk{f}}{\dot }^{%
2}+m^{2}\SGMPmathgrk{f}^{2}}$\cr
$\displaystyle{{\left\LeftPost{par}\MthAcnt {{{a}\over{a}}}{\dot }
\right\RightPost{par}}^{{\bf \cdot~}}+{\left\LeftPost{par}\MthAcnt {%
{{a}\over{a}}}{\dot }\right\RightPost{par}}^{%
2}=-2\MthAcnt {\SGMPmathgrk{f}}{\dot }^{2}+m^{%
2}\SGMPmathgrk{f}^{2}}$\cr
}}}$\cr
}}\end{equation}

\par

\par Eqs. (51) are invariant under the transformation
\(
t{\ifmmode\rightarrow\else$\rightarrow$\fi}-t\).
The three equations of motion (51) can be combined
in one equation, in which the time parameter
\( t\) does not appear at all:
\def\XRefId{}
\begin{equation}\SGMPlab\XRefId\vcenter{\halign{\strut\hfil#\hfil&#\hfil\cr
$\displaystyle{\SGMPmathgrk{f}{{d^{2}\SGMPmathgrk{f}}\over{d\SGMPmathgrk{a}^{%
2}}}+{\left\LeftPost{par}3\SGMPmathgrk{f}{{d\SGMPmathgrk{f}}
\over{d\SGMPmathgrk{a}}}+1\right\RightPost{par}}{\left\LeftPost{sqb}1
-{\left\LeftPost{par}{{d\SGMPmathgrk{f}}\over{d\SGMPmathgrk{a}}}
\right\RightPost{par}}^{%
2}\right\RightPost{sqb}}=0}$\cr
}}\end{equation}

(For convenience, we use the variable $\alpha$ = 1n {\em a}
here and below.) This equation completely describes the classical
trajectories in the
\(
{\left\LeftPost{par}\SGMPmathgrk{a},\hskip 0.167em
\SGMPmathgrk{f}\right\RightPost{par}}
\)
space. The direction of motion along the trajectories
is determined by the choice of the time direction.\par

\par At this point we should comment on whether or not the sign of the
action S has anything to do with the expansion or contraction of a
cosmological model. This issue is often discussed in the context of the
so-called {``}tunneling{''} wave function [22]. Eqs.~(48) and (50)
may lead to the impression that
\(
S{\ifmmode<\else$<$\fi}0\hskip 0.167em {\left\LeftPost{par}f
{\ifmmode>\else$>$\fi}0\right\RightPost{par}}
\)
corresponds to the expansion
\(
{\left\LeftPost{par}\MthAcnt {a}{\dot }{\ifmmode>\else$>$\fi}0
\right\RightPost{par}}
\),
while the opposite choice
\(
S{\ifmmode>\else$>$\fi}0\hskip 0.167em {\left\LeftPost{par}f
{\ifmmode<\else$<$\fi}0\right\RightPost{par}}
\)
corresponds to the contraction
\(
{\left\LeftPost{par}\MthAcnt {a}{\dot }{\ifmmode<\else$<$\fi}0
\right\RightPost{par}}
\)
of the cosmological volume. However, the choice of the parameter
\( t\) in these equations is absolutely arbitrary. The
functions
\(
S{\ifmmode>\else$>$\fi}0\hskip 0.167em {\left\LeftPost{par}f
{\ifmmode<\else$<$\fi}0\right\RightPost{par}}
\)
can perfectly well describe expansion if one changes
the parameter
\( t\) to \( -t\)
in Eq. (50). Thus, the sign of the action does not
prescribe a particular meaning to the direction of evolution along the
classical trajectories.\par

\par For the model defined by classical equations of motion (51), all
trajectories of the model in the
\(
{\left\LeftPost{par}\SGMPmathgrk{f},\hskip 0.167em \MthAcnt
{\SGMPmathgrk{f}}{\dot }
\right\RightPost{par}}\)
 phase plane have previously been found [23]. It has
been shown that in the case of expansion (i.e., \.a {\ifmmode>\else$>$\fi} 0),
all the
trajectories, except for two, start out from the ejecting nodes
\( K_{1}\) and
\( K_{2}\) (see Fig. 6). The remaining two trajectories,
corresponsing to the inflationary regime, are two attracting
separatrices that originate at the saddle points
\( S_{1}\) and
\( S_{2}\). The solutions to Eq. (49) have the following
asymptotic behavior for trajectories that start out from the nodes:
\(
f{\ifmmode\approx\else$\approx$\fi}ce^{{\ifmmode\pm\else$\pm$\fi}3\SGMPmathgrk{f}},\hskip 0.265em c^{2}
e^{{\ifmmode\pm\else$\pm$\fi}3\SGMPmathgrk{f}}
{\ifmmode\gg\else$\gg$\fi}m^{2}\SGMPmathgrk{f}^{2}
,\hskip 0.265em c=const\).
And for the separatrices one has
\(
f{\ifmmode\approx\else$\approx$\fi}{\ifmmode\pm\else$\pm$\fi}
{{1}\over{3}}m\SGMPmathgrk{f},\hskip 0.265em
9\SGMPmathgrk{f}^{2}{\ifmmode\gg\else$\gg$\fi}1.\)
Different values of the constant
\( c\) select different trajectories leaving the nodes.
Hence, a particular solution to Eq. (49) gives a difinite function
\( S\) and, at the same time, a particular classical
trajectory.\par

\par Now we will relate different wave functions to different classical
solutions. One can distinguish different solutions to Eq.~(49) by the
subscript
\( n\) which varies continuously and takes on two distinct
values corresponding to the separatrices. By virtue of the linearity of
the WD equation, we can present a full set of solutions to Eq.~(45) in
the form
\(
\SGMPmathgrk{y}=\sum _{n}exp{\left\LeftPost{par}iA_{%
n}+iS_{n}\right\RightPost{par}},\hskip 0.265em S_{%
n}=-exp{\left\LeftPost{par}3\SGMPmathgrk{a}\right\RightPost{par}}f_{%
n},\hskip 0.265em A_{n}=const\)
valid in the lowest approximation. To every
quasi-classical wavefunction $\psi$${}_{n}$ =
exp({\em iS${}_{n}$})  one can put into
correspondence a family of lines that are ortogonal to the surfaces
{\em S${}_{n}$} = const (Fig. 7). These surfaces are
constructed in the minisuperspace
\(
{\left\LeftPost{par}\SGMPmathgrk{a},\hskip 0.167em
\SGMPmathgrk{f}\right\RightPost{par}}
\)
endowed with the metric tensor
\def\XRefId{}
\begin{equation}\SGMPlab\XRefId\vcenter{\halign{\strut\hfil#\hfil&#\hfil\cr
$\displaystyle{G^{\SGMPmathgrk{m}\SGMPmathgrk{n}}=
e^{-3\SGMPmathgrk{a}}diag{\left\LeftPost{par}
-1,\hskip 0.212em +1\right\RightPost{par}},\hskip 0.265em \hskip 0.265em
\hskip 0.212em \hskip 0.265em \hskip 0.265em \SGMPmathgrk{m},\SGMPmathgrk{n}
=1,2,\hskip 0.265em \hskip 0.265em \hskip 0.265em \hskip 0.265em
\hskip 0.265em \hskip 0.265em x^{1}=\SGMPmathgrk{a},\hskip 0.265em
\hskip 0.265em \hskip 0.265em x^{2}=\SGMPmathgrk{f}}$\cr
}}\end{equation}

The vector
\(
N_{\SGMPmathgrk{m}}\)
 ortogonal to
\(
S_{n}\)
 = const can be obtained by acting on
$\psi$${}_{n}$ =  exp({\em iS${}_{n}$}) with the
momentum operators $\pi{}_{\alpha}$ and
$\pi{}_{\phi}$:\SGMPnewline
\SGMPnewline
 ~~~~~~~~~~~~~~~~~~~~~~~~~~~~~
\(
\vcenter{\halign{\strut\hfil#\hfil&#\hfil\cr
$\displaystyle{\MthAcnt {\SGMPmathgrk{p}}{\hat }_{\SGMPmathgrk{a}
}\SGMPmathgrk{y}_{n}={{1}\over{i}}\hskip 0.167em
{{\SGMPmathgrk{6}}\over{\SGMPmathgrk{6}\SGMPmathgrk{a}}}\hskip 0.167em
\SGMPmathgrk{y}_{n}={{\SGMPmathgrk{6}S_{n}}\over{\SGMPmathgrk{6}
\SGMPmathgrk{a}}}\SGMPmathgrk{y}_{n}=N_{\SGMPmathgrk{a}}\SGMPmathgrk{y}_{%
n}}$\cr
$\displaystyle{\MthAcnt {\SGMPmathgrk{p}}{\hat }_{\SGMPmathgrk{f}}
\SGMPmathgrk{y}_{n}={{1}\over{i}}\hskip 0.167em
{{\SGMPmathgrk{6}}\over{\SGMPmathgrk{6}\SGMPmathgrk{f}}}\hskip 0.167em
\SGMPmathgrk{y}_{n}={{\SGMPmathgrk{6}S_{n}}\over{\SGMPmathgrk{6}
\SGMPmathgrk{f}}}\SGMPmathgrk{y}_{n}=N_{\SGMPmathgrk{f}}\SGMPmathgrk{y}_{%
n}}$\cr
}}\)
\SGMPnewline
\SGMPnewline
Taking into account Eqs. (48), (53) one obtains N${}^{\alpha}$ =
3f, N${}^{\phi}$ = {--}f${}^{{\ifmmode{}^\prime\else${}^\prime$\fi}}$.
The vector
field
\(
{\left\LeftPost{par}N^{\SGMPmathgrk{a}},
\hskip 0.167em N^{\SGMPmathgrk{f}
}\right\RightPost{par}}\)
 determines the lines
\(
x^{\SGMPmathgrk{m}}{\left\LeftPost{par}\SGMPmathgrk{a},
\hskip 0.167em \SGMPmathgrk{f}
\right\RightPost{par}}\)
ortogonal to
\( S_{n}={\rm const}\) in a parametric way:
\(
dx^{\SGMPmathgrk{m}}/dt=N^{\SGMPmathgrk{m}}\).
These lines coincide with the classical trajectories
(see Eq. (50)). One can also note that by integrating the
relation~
\(
d\SGMPmathgrk{a}/d\SGMPmathgrk{f}=N^{\SGMPmathgrk{a}}/N^{\SGMPmathgrk{f}}
=-3f/f'\)
along every classical path in the ($\alpha$, $\phi$) plane
one gets z($\alpha$, $\phi$) = const where
\(
z\equiv \SGMPmathgrk{a}+3\int {\left\LeftPost{par}f/f'\right\RightPost{par}}
d\SGMPmathgrk{f}\).\par

\par In the case at hand, the family of lines ortogonal to S${}_{n}$
and the associated tangent vectors N${}^{\alpha}$,
N${}^{\phi}$ are independent of $\alpha$ and transform into
themselves under the shift $\alpha$
{\ifmmode\rightarrow\else$\rightarrow$\fi} $\alpha$+ const, or a(t)
{\ifmmode\rightarrow\else$\rightarrow$\fi}const a(t).
This symmetry is a reflection of the fact that the
function a(t) alone does not appear in Eq.~(51), it appears only as the
Hubble factor
\( \MthAcnt {a}{\dot }/a\).
Therefore, the invariance of the vector field
\( N^{\SGMPmathgrk{m}}\) under the displacement $\alpha$
{\ifmmode\rightarrow\else$\rightarrow$\fi} $\alpha$ + const
means that, for a given {\em S${}_{n}$}, the lines
traced out by
\( N^{\SGMPmathgrk{m}}\) are all copies of one and the same physically
distinct classical solution. It happened as a consequence of our
assumption of a negligibly small spatial curvature, k=0; in general, it
is not the case.\par

\par Thus, we see that different solutions to the Hamilton-Jacobi
equation determine different wave functions in their lowest (in terms of
\(
{\ifmmode\hbar\else$\hbar$\fi}\)
) approximation. On the other hand, to a given
{\em S${}_{n}$} one can assign a family of classical
trajectories. The next approximation to
{\em S${}_{n}$} defines the prefactor to the wave
function
\( \SGMPmathgrk{y}_{n}=e^{iSn}\).
The prefactor is responsible for forming a packet
from classical trajectories determined by
{\em S${}_{n}$}. It assigns different
{``}weights{''} to different classical paths.\par

\par Let us return to Eq. (47) for
{\em $\sigma$${}_{n}$}. The general solution for
{\em $\sigma$${}_{n}$} can be expressed in terms of the
function f${}_{n}$($\phi$):
\(
\SGMPmathgrk{s}_{n}{\left\LeftPost{par}\SGMPmathgrk{a},
\SGMPmathgrk{f}\right\RightPost{par}}
={{i}\over{2}}{\left\LeftPost{par}3\SGMPmathgrk{a}+ln\hskip 0.212em
f^{{\ifmmode{}^\prime\else${}^\prime$\fi}}_{n}\right\RightPost{par}}+B_{n}
{\left\LeftPost{par}z\right\RightPost{par}}\)
where B${}_{n}$ is an arbitrary function of its
argument z. In the considered approximation, the general solution to WD
equation can be written in the form
\def\XRefId{}
\begin{equation}\SGMPlab\XRefId\vcenter{\halign{\strut\hfil#\hfil&#\hfil\cr
$\displaystyle{\SGMPmathgrk{y}=\sum _{n}Z_{n}\SGMPmathgrk{y}_{%
n}=\sum _{n}Z_{n}exp{\left\LeftPost{par}
iS_{n}+i\SGMPmathgrk{s}_{n}\right\RightPost{par}}}$\cr
}}\end{equation}
where
\def\XRefId{}
\begin{equation}\SGMPlab\XRefId\vcenter{\halign{\strut\hfil#\hfil&#\hfil\cr
$\displaystyle{\SGMPmathgrk{y}_{n}=\SGMPmathgrk{c}_{n}e^{i\SGMPmathgrk{g}_{%
n}}{\left\LeftPost{par}a^{3}f^{{\ifmmode{}^\prime\else${}^\prime$\fi}}_{%
n}\right\RightPost{par}}^{-1/2}e^{-ia^{3}
f_{n}},}$\cr
}}\end{equation}

\noindent
$\chi$${}_{n}$ and $\gamma$${}_{n}$ are arbitrary real functions of
z and Z${}_{n}$ are arbitrary complex numbers. One can see that to
every path z = const in ($\alpha$, $\phi$) plane one can assign a number
Q${}_{n}$
\(
\equiv \)
$\chi$${}_{n}$${}^{2}$ (z) which is
conserved along this path. A particular value of Q${}_{n}$ is
determined by a chosen wave function (in other words, by the chosen
boundary conditions for the wave function) and, specifically by the
function $\chi$${}_{n}$(z). Diffrent wave functions favor the
inflationary trajectories to a different degree (see, for example,
[24]).\SGMPnewline
\par

\par \underline{{\protect\normalsize {\bf 9.
 From the Space of Classical Solutions to the Space of}}}\newline
\underline{{{\bf Wave
Functions}}}\par

\par From the problem of distributing {``}weights{''} among
different classical trajectories belonging to the same family determined
by {\em S${}_{n}$} we now turn to the more difficult
problem of distributing {``}weights{''} among the wavefunctions
themselves. As we saw above, the WKB components $\psi$ ${}_{n}$,
Eq.~(55), participate in the general solution Eq.~(54) with arbitrary
complex coefficients Z${}_{n}$. They determine one or other choice
of possible wavefunctions. How can one classify the space of all
possible wavefunctions? \par

\par To answer this question we will start from the simplest situation,
when the number of the linearly independent solutions to WD equation is
only two. For this aim we will first consider Eq.~(44) in another
limiting case, namely, when the term
\(
-{{1}\over{a^{2}}}\hskip 0.167em {{%
\SGMPmathgrk{6}^{2}}\over{\SGMPmathgrk{6}\SGMPmathgrk{f}^{2}}}
\)
can be neglected. In this case the variable $\phi$
plays the role of a parameter and the problem reduces to a
one-dimensional problem. The basic Eq.~(44) can be written down in the
form (for k = +1):
\def\XRefId{}
\begin{equation}\SGMPlab\XRefId\vcenter{\halign{\strut\hfil#\hfil&#\hfil\cr
$\displaystyle{{\left\LeftPost{par}{{1}\over{a^{p}}}\hskip 0.167em
{{d}\over{da}}\hskip 0.167em a^{p}{{%
d}\over{da}}-a^{2}+H^{2}a^{4}
\hskip 0.167em \right\RightPost{par}}
\SGMPmathgrk{y}{\left\LeftPost{par}a\right\RightPost{par}}
=0}$\cr
}}\end{equation}

\noindent
where
\( H\equiv m^{2}\SGMPmathgrk{f}^{2}\).
We prefer to work with exact solutions to Eq.~(56)
so we choose p = {--} 1 or p = 3 [25]. (The case p = {--} 1 was
first considered in Ref. [26].) We will write the exact solution for the
p = {--} 1 case in the form:
\def\XRefId{}
\begin{equation}\SGMPlab\XRefId\vcenter{\halign{\strut\hfil#\hfil&#\hfil\cr
$\displaystyle{\vcenter{\halign{\strut\hfil#\hfil&#\hfil\cr
 $\displaystyle{\SGMPmathgrk{y}{\left\LeftPost{par}a\right\RightPost{par}}=u^{%
1/2}{\left\LeftPost{sqb}A_{1}H^{{\left\LeftPost{par}1\right\RightPost{par}}
}_{1/3}{\left\LeftPost{par}{{u^{3/2}}
\over{3H^{2}}}\right\RightPost{par}}+A_{2}
H^{{\left\LeftPost{par}2\right\RightPost{par}}}_{1/3}
{\left\LeftPost{par}{{u^{3/2}}\over{3H^{2}}}
\right\RightPost{par}}\right\RightPost{sqb}},\hskip 0.265em \hskip 0.265em
\hskip 0.265em H^{2}a^{2}{\ifmmode\geq\else$\geq$\fi}1,}$\cr
$\displaystyle{}$\cr
}}\hskip 0.265em \hskip 0.265em \hskip 0.265em \hskip 0.265em
}$\cr
}}\end{equation}

\def\XRefId{}
\begin{equation}\SGMPlab\XRefId\vcenter{\halign{\strut\hfil#\hfil&#\hfil\cr
$\displaystyle{\SGMPmathgrk{y}{\left\LeftPost{par}
a\right\RightPost{par}}={\left\LeftPost{par}
-u\right\RightPost{par}}^{1/2}{\left\LeftPost{sqb}B_{1}
I_{1/3}{\left\LeftPost{par}{{{\left\LeftPost{par}-
u\right\RightPost{par}}^{%
3/2}}\over{3H^{2}}}\right\RightPost{par}}
+B_{2}K_{1/3}{\left\LeftPost{par}{{{\left\LeftPost{par}
-u\right\RightPost{par}}^{3/2}}\over{3H^{2}}}
\right\RightPost{par}}\right\RightPost{sqb}},\hskip 0.265em \hskip 0.265em
H^{2}a^{2}{\ifmmode\leq\else$\leq$\fi}1}$\cr
}}\end{equation}

\noindent
where $u = H^{2}a{}^{2} -1$ and
I${}_{1/3}$, K${}_{1/3}$, H${}^{(1)}$${}_{1/3}$,
H${}^{(2)}$${}_{1/3}$ are the Infeld, Macdonald and Hankel
special functions, correspondingly.\par

\par Eq. (56) has the form of the Schr\"odinger equation for a
1{--}dimensional problem with the potential V(a) = a${}^{2}$
{--} H${}^{2}$a${}^{4}$ (see Fig. 8). The
coefficients A${}_{1}$, A${}_{2}$ and B${}_{1}$,
B${}_{2}$ are two pairs of arbitrary constant coefficients in front
of two pairs of linearly independent solutions. By matching the
solutions (57) and (58) at the point a = 1/H one finds [25]:
\(
B_{1}=-A_{1}{\left\LeftPost{par}1+i\Rad{3}\DoRad
\right\RightPost{par}}-A_{2}{\left\LeftPost{par}1-i\Rad{
3}\DoRad \right\RightPost{par}},\hskip 0.265em \hskip 0.265em
\hskip 0.265em B_{2}={{2i}\over{\SGMPmathgrk{p}}}
{\left\LeftPost{par}A_{2}-A_{1}\right\RightPost{par}}
\).\par

\par Now let us characterize the full space of the wave functions. (Here
we mainly follow Ref. [27].) In the present case, our quantun system has
only two linearly independent states and therefore resembles a simple
spin - 1/2 system. Let us call an arbitrarily chosen basis of states
\( |1>\) and
\( |2>\). A general state
\( |\SGMPmathgrk{y}>\)
can be expanded as {\ifmmode|\else$|$\fi}$\psi${\ifmmode>\else$>$\fi} =
 Z${}_{1}${\ifmmode|\else$|$\fi}1{\ifmmode>\else$>$\fi} +
Z${}_{2}$ {\ifmmode|\else$|$\fi} 2 {\ifmmode>\else$>$\fi},
where Z${}_{1}$ and Z${}_{2}$ are arbitrary
complex constants.\par

\par It is a general principle of quantum mechanics that state vectors
which differ only by an overall non-zero multiple $\lambda$ describe one and
the same physical state. Thus, the pair of coordinates (Z${}_{1}$
and Z${}_{2}$) and the pair ( $\lambda$Z${}_{1}$ and
$\lambda$Z${}_{2}$) are equivalent. It follows that physical quantities
can only depend on the ratio $\zeta$ = Z${}_{1}$/Z${}_{2}$ which
is invariant under rescaling. In our example above we may identify
Z${}_{1}$ with B${}_{1}$ and Z${}_{2}$ with B${}_{2}$.
It is convenient to introduce the notation
\(
B_{1}={\left\LeftPost{vb}B_{1}\right\RightPost{vb}}e
xp{\left\LeftPost{par}i\SGMPmathgrk{b}_{1}\right\RightPost{par}},\hskip 0.265em
\hskip 0.265em \hskip 0.265em B_{2}={\left\LeftPost{vb}B_{%
2}\right\RightPost{vb}}exp{\left\LeftPost{par}i\SGMPmathgrk{b}_{2}
\right\RightPost{par}},\hskip 0.212em \hskip 0.265em \hskip 0.265em
\SGMPmathgrk{b}=\SGMPmathgrk{b}_{1}-\SGMPmathgrk{b}_{2}\)
and then
\(
\SGMPmathgrk{z}=B_{1}/B_{2}=x\hskip 0.167em exp{\left\LeftPost{par}
i\SGMPmathgrk{b}\right\RightPost{par}}.\)
The ratio $\zeta$ parameterizes the points on a 2
{--} dimensional sphere and so we see that the set of possible
wavefunctions is in 1{--}1 correspondence with the points on the 2
{--} sphere. \par

\par We now wish to place a measure on the space of quantum states. Of
course there are many possible measures. However, in choosing a measure
we should be guided by the priniciple that the measure should be
independent of the arbitrary choice of basis states
\( |1>\) and \( |2>\).
That is if we perform a unitary change of basis,
which will preserve all probability amplitudes, then the measure should
remain invariant.\par

\par The invariance of the measure may be taken as the quantum analogue
of the principle of general covariance in classical general relativity.
In fact in the classical limit it corresponds to invariance under
canonical transformations. This latter invariance was used in Ref. [28]
to suggest a suitable measure on the set of classical solutions. \par

\par For a 2{--}state system the 2{--}dimensional unitary
transformations will act (provided
\( |1>\) and \( |2>\) are normalised) on the complex 2{--}vector
(Z${}_{1}$ and Z${}_{2}$) by multiplication by a 2 by 2 unitary
matrix. Clearly the ratio $\zeta$ is unaffected by matrices which are
merely multiples of the unit matrix so we may confine attention to
special unitary matrices of determinant unity, this still allows minus
the identity matrix so if we want just the transformations which change
the physical states we must identify to SU(2) matrices which differ by
multiplication by minus one. That is, the effective physical
transformations acting on the space of quantum states is the rotation
group SO(3) = SU(2)/C${}_{2}$ where C${}_{2}$ is the group
consisting of + 1 and {--} 1. In fact this acts on the 2{--}sphere
in the usual way provided we identify $\beta$ with the longitudinal angle
and x = cotan($\theta$/2) where $\theta$ is the usual co-latitude.\par

\par It is now clear that we must choose for our invariant measure on
the space of quantum states the usual volume element on the
2{--}sphere. This is clearly invariant under rotations and up to an
arbitrary constant multiple it is unique. That is the measure in terms
of $\beta$ and $\theta$ is:
\def\XRefId{}
\begin{equation}\SGMPlab\XRefId\vcenter{\halign{\strut\hfil#\hfil&#\hfil\cr
$\displaystyle{dV=\sin \SGMPmathgrk{q}d\SGMPmathgrk{q}d\SGMPmathgrk{b},\hskip
0.265em
\hskip 0.265em \hskip 0.265em \hskip 0.265em \hskip 0.265em
0{\ifmmode\leq\else$\leq$\fi}\SGMPmathgrk{q}
{\ifmmode\leq\else$\leq$\fi}\SGMPmathgrk{p},
\hskip 0.265em \hskip 0.265em \hskip 0.265em
\hskip 0.265em \hskip 0.265em 0
{\ifmmode\leq\else$\leq$\fi}\SGMPmathgrk{b}
{\ifmmode\leq\else$\leq$\fi}2\SGMPmathgrk{p}}$\cr
}}\end{equation}

\noindent
Of course the measure is just the Riemannian volume element with respect
to the standard round metric on the 2{--}sphere.\par

\par It should be mentioned that the well known Hartle-Hawking
wavefunction [29] is exactly the south pole ($\theta$ = $\pi$) of the
2{--}sphere. This wavefunction is real. Another real wavefunction
corresponds to the north pole of the 2{--}sphere. We call this
wavefunction anti-Hartle-Hawking wavefunction. All other wavefunctions
are complex.\SGMPnewline
\par

\par \underline{{\protect\normalsize {\bf 10.
On the Probability of Quantum Tunneling from
{``}Nothing{''}}}}\par

\par The measure introduced in the space of all wave-functions may allow
us to formulate and solve some physically meaningful problems. We will
try to pose one such a problem already in the considered simplest model.
As was mentioned above, Eq. (56) looks like the Schr\"odinger
equation for a particle of a zero energy moving in the presence of the
potential V(a). The form of the potential (Fig. 8) motivates the
expectation that some of the wave-functions may be capable of describing
the quantum tunneling or decay. In ordinary quantum mechanics the
quantity D, where~
\(
D={\left\LeftPost{vb}{{\SGMPmathgrk{y}
{\left\LeftPost{par}a_{2}\right\RightPost{par}}
}\over{\SGMPmathgrk{y}{\left\LeftPost{par}a_{1}\right\RightPost{par}}
}}\right\RightPost{vb}}^{2}\),
can be interpreted as the quasiclassical probability
for the particle to tunnel from one classically allowed region to
another (see Fig. 8). The wave function used in this expression is
determined by the imposed boundary conditions, i.e. it is determined by
the physical formulation of the problem. The value of D is always (much)
less than unity, D{\ifmmode<\else$<$\fi}1, for
wave functions describing quantum tunneling
or decay. One can define a similar quantity D in our quantum
cosmological model, though, the physical interpretation of D is less
clear. The main difference is that in ordinary quantum mechanics one
imposes suitable boundary conditions in time t and space x, while in our
problem there is only one coordinate, a. (The notion of the break of
classical evolution in quantum cosmology is rather delicate. We have
argued in Ref. [30], that only in superspaces of more than one
dimension, this notion can be clearly formulated.) Nevertheless, we will
adopt the same definition of D in our problem:
\(
D={\left\LeftPost{vb}{{\SGMPmathgrk{y}{\left\LeftPost{par}{{1}\over{%
H}}\right\RightPost{par}}}\over{\SGMPmathgrk{y}{\left\LeftPost{par}
0\right\RightPost{par}}}}\right\RightPost{vb}}^{2} \),
and will consider
\( p{\ifmmode<\else$<$\fi}1\). The quantity
\( D\)
is well defined mathematically and can be calculated
for every solution of Eq. (56) regardless of its interpretation. Since
in quantum{--}mechanical problem the {``}energy{''}
\( \SGMPmathgrk{e}\)
of the particle is $\epsilon$ = 0, we can provisionally
interpret D as the probability for creation of the universe from
{``}nothing{''}. (This is more precise formulation of the notion,
introduced at the beginning of these lectures and graphically depicted
in Fig. 1.) Therefore, we are interested in dividing the wave functions
into two classes which predict D{\ifmmode<\else$<$\fi}1
and D{\ifmmode>\else$>$\fi}1. It is not excluded that
the wave-functions predicting D{\ifmmode<\else$<$\fi}1
can be eventually justified as
describing the quantum tunneling from {``}nothing{''}, or rather
from the {``}vacuum{''} defined in the framework of some more deep
quantum theory.\par

\par It is easy to calculate D in the approximation H
{\ifmmode<\else$<$\fi}{\ifmmode<\else$<$\fi} 1 [25].
One can see that the different choices of the wave-function give
different values of D. For instance, the Hartle-Hawking wave-function
corresponds to the choice B${}_{1}$=0 and gives
\(
D=exp{\left\LeftPost{par}{{2}\over{3H^{2}}}
\right\RightPost{par}}{\ifmmode\gg\else$\gg$\fi}1\).
We are interested to know the value of D for a
typical wave-function. In other words, we need to know how many
wave-functions give D{\ifmmode<\else$<$\fi}1 or
D{\ifmmode>\else$>$\fi}1? To answer this question one must
consider the space of all possible wave-functions with a suitable
measure. By using the measure (59) one can show that the set of wave
functions predicting D{\ifmmode>\else$>$\fi}1 is very small compared
with that predicting
D{\ifmmode<\else$<$\fi}1. This follows from the fact that the surface
area of the patch
covered by the wave-functions with D{\ifmmode>\else$>$\fi}1 is very
small compared with the
total surface area of the 2{--}sphere. Indeed, the circle separating
D{\ifmmode>\else$>$\fi}1 and D{\ifmmode<\else$<$\fi}1 regions on
the 2{--}sphere corresponds to the value
\(
\SGMPmathgrk{q}_{0}{\ifmmode\approx\else$\approx$\fi}\SGMPmathgrk{p}-
2\hskip 0.265em exp{\left\LeftPost{par}
-{{1}\over{3H^{2}}}\right\RightPost{par}}
,\hskip 0.265em \hskip 0.265em \hskip 0.265em \hskip 0.265em
\hskip 0.265em
\SGMPmathgrk{p}-\SGMPmathgrk{q}_{0}{\ifmmode\ll\else$\ll$\fi}1.\)
Only a small area around the south pole $\theta$ =
$\pi$ gives the wave-functions with D{\ifmmode>\else$>$\fi}1, the rest
of the surface of the
2{--}sphere corresponds to the wave-functions with
D{\ifmmode<\else$<$\fi}1. The ratio
of the surface area around the south pole to the total surface area is
very small; it is equal to
\(
exp{\left\LeftPost{par}-{{2}\over{3H^{2}}}\right\RightPost{par}}
{\ifmmode\ll\else$\ll$\fi}1.\)
Thus, one can say, that the probability of finding a
wave function with D{\ifmmode>\else$>$\fi}1 (among them is the
Hartle-Hawking wave-function)
is very small. One can conclude that the overwhelming majority of the
wave-functions seem to be capable of describing the quantum tunneling or
decay, since they predict D{\ifmmode<\else$<$\fi}1. (It is interesting
to note that the
product of surface areas with D{\ifmmode>\else$>$\fi}1 and
D{\ifmmode<\else$<$\fi}1 to their corresponding
maximal values of D gives approximately equal numbers, both of order
unity.)\par

\par The simple example presented above clarifies the notion of the
measure in the space of all physically distinct wave-functions. In a
similar way one can introduce the measure in the multidimensional space
of the wave-functions described by Eq.~(54) [27]. The use of this
measure shows that the inflation is indeed a property of a typical wave
function, at least, under some additional assumptions adopted in [27,
25].\SGMPnewline
\par

\par \underline{{\protect\normalsize {\bf 11.
Duration of Inflation and Possible Remnants of the}}}\newline
\underline{{{\bf Preinflationary
Universe}}}\par

\par In the framework of the inflationary hypothesis, one normally
considers cosmological models whose period of inflation lasted much
longer than the minimal duration necessary to increase a preinflationary
scale to the size of the present-day Hubble radius. In such models, the
number
\( N\) of \( e\)-foldings of the scale factor during inflation is much
larger than the minimal
\( N_{min}\), in which case the volume covered by inflation is
much larger than the present-day Hubble volume. For this reason, one
normally does not expect to find any {``}remnants{''} of the
preinflationary universe (see, however, Ref. [31]) as they were
enormously diluted and spread over the huge inflated volume.
Nevertheless, according to the quantum cosmological considerations, the
duration of inflation close to the minimally sufficient amount may
happen to be the most probable prediction of some popular quantum
cosmological models, as we will see below. (This section is based on
Ref. [32]).\par

\par Quantum cosmology is supposed to provide initial data for classical
cosmological models and resolve such issues as the likelihood of
inflation and its probable duration. Obviously, we are still far away
from a satisfactory answer. A part of the problem is that there are too
many possible wave functions: the trouble of selecting an appropriate
classical solution from the space of all possible classical solutions is
replaced by an even bigger problem of selecting an appropriate wave
function from the space of all possible wave functions. However, if a
cosmological wave function is chosen, the derivation of the probability
distribution of the permitted classical solutions seems to be more
straightforward.\par

\par A wave function which has received much attention in the literature
is the Hartle-Hawking wave function
\( \SGMPmathgrk{y}_{HH}\).
As we have seen above, one cannot say that the
\( \SGMPmathgrk{y}_{HH}\)
is in any sense more probable than others. On the
contrary, it looks, rather, as an exception. For simple quantum
cosmological models allowing inflation, the Hartle-Hawking wave function
corresponds to a single point {---} a pole on the two-sphere
representing the space of all physically different wave functions (see
Sec.~9). However, the
\( \SGMPmathgrk{y}_{HH}\)
is a real wave function while all others (except the
one corresponding to the opposite pole which is also real and which we
call the {``}anti-Hartle-Hawking{''} wave function) are complex.
This exceptional property of the
\( \SGMPmathgrk{y}_{HH}\)
alone, if for no other reasons, justifies special
attention to this wave function and makes it interesting to see what
kind of predictions with regard to inflation follow from it.\par

\par For the case of homogeneous isotropic models with the scale factor
\(
a{\left\LeftPost{par}t\right\RightPost{par}}\)
and a scalar field
\(
\SGMPmathgrk{f}{\left\LeftPost{par}t\right\RightPost{par}}\),
the
\( \SGMPmathgrk{y}_{HH}\)
predicts a set of classical inflationary solutions
which can be described as trajectories in the two-dimensional space
\(
{\left\LeftPost{sqb}a{\left\LeftPost{par}t\right\RightPost{par}},\hskip 0.167em
\SGMPmathgrk{f}{\left\LeftPost{par}t\right\RightPost{par}}\right\RightPost{sqb}}
\)
(see [21] and Sec.~8). These trajectories begin in
the vicinity of a line which is the caustic line for the so-called
Euclidean trajectories. The probability distribution
\( P_{HH}\) for the classical (Lorentzian) inflationary solutions
follows from the
\( \SGMPmathgrk{y}_{HH}\) and has the form
\def\XRefId{}
\begin{equation}\SGMPlab\XRefId\vcenter{\halign{\strut\hfil#\hfil&#\hfil\cr
$\displaystyle{P_{HH}=N\hskip 0.167em {\rm exp}{\left\LeftPost{sqb}
{{2}\over{3H^{2}{\left\LeftPost{par}\SGMPmathgrk{f}\right\RightPost{par}}
}}\right\RightPost{sqb}},}$\cr
}}\end{equation}

\noindent
where
\( N\) is the normalization constant and
\(
H{\left\LeftPost{par}\SGMPmathgrk{f}\right\RightPost{par}}\)
is the Hubble factor at the beginning of inflation.
The function
\( P_{HH}\) varies along the caustic line and increases rapidly
toward the smaller values of
\( \SGMPmathgrk{f}\).
This means that the probability to find a given
inflationary solution is higher the lower the initial value of the
scalar field
\(
\SGMPmathgrk{f}{\left\LeftPost{par}t\right\RightPost{par}}\)
(if, of course, this interpretation of
\( P_{HH}\) is correct). But smaller initial values of
\(
\SGMPmathgrk{f}{\left\LeftPost{par}t\right\RightPost{par}}\)
correspond to the shorter periods of inflation which
makes solutions with a shorter period of inflation much more probable
than solutions with a longer period of inflation.\par

\par An important fact is, however, that the inflationary period cannot
be too short. The reason is that the caustic line does not extend down
to the very low values of
\( \SGMPmathgrk{f}\);
instead, it has a sharp cusp (singularity) at the
point of return from which the second branch of the caustic line
develops (see Fig. 9) [24]. The point of return on the caustic line
divides the Euclidean trajectories into two families which touch the
first or the second branch of the caustic, respectively. The Lorentzian
inflationary solutions cannot begin with the initial value of the scalar
field and the Hubble factor lower than the value corresponding to the
point of return
\( \SGMPmathgrk{f}^{*}\)
and, therefore, their periods of inflation cannot be
arbitrarily short. Thus,
\( \SGMPmathgrk{y}_{HH}\)
gives more weight to inflationary solutions with lower initial values of
\(
\SGMPmathgrk{f}{\left\LeftPost{par}t\right\RightPost{par}}\)
but does not accommodate solutions which begin with
\(
\SGMPmathgrk{f}{\left\LeftPost{par}t\right\RightPost{par}}\)
smaller than
\(
\SGMPmathgrk{f}^{*}\).
The numerical estimates for the case of the scalar field potentials
\(
V{\left\LeftPost{par}\SGMPmathgrk{f}\right\RightPost{par}}=
m^{2}\SGMPmathgrk{f}^{%
2}/2\)
and
\(
V{\left\LeftPost{par}\SGMPmathgrk{f}\right\RightPost{par}}=
\SGMPmathgrk{l}\SGMPmathgrk{f}^{%
4}/2\)
show [24] that the number
\( \SGMPmathgrk{f}^{*}\)
falls short a factor 4 or 3, respectively, to ensure
the minimally sufficient inflation. The inflated scale turns out to be
of order 10${}^{21}$ cm instead of the required
10${}^{28}$ cm. At the same time, the probability distribution
function
\( P_{HH}\)
reaches its maximum value at
\(
\SGMPmathgrk{f}=\SGMPmathgrk{f}^{*}\).
Thus, it seems that the most probable prediction of
the Hartle-Hawking wave function is a {``}small, underinflated
universe{''}. However, it is possible that the discrepancy between
\( l_{H}\)
and the predicted inflated scale may be weakened or
even removed for other scalar field potentials. Apart from that, the
deficiency of
\(
\SGMPmathgrk{f}^{*}\)
in being just a numerical factor 4 or 3 smaller than
necessary, in the situation where the initial values of the scalar field
can vary within a huge interval from
\( \SGMPmathgrk{f}^{*}\)
up to about 10${}^{5}$
\( \SGMPmathgrk{f}^{*}\),
can serve as an indication that the duration of
inflation close to the minimally sufficient amount should, probably, be
taken seriously, at least, as a prediction of the Hartle-Hawking wave
function.\par

\par The meaning of the above discussion is that the search for the
{``}remnants{''} of the preinflationary universe, in the framework
of the inflationary hypothesis, may not necessarily be of a purely
academic interest.\SGMPnewline
\par

\par \underline{{\protect\normalsize {\bf 12.
Relic Gravitons and the Birth of the
Universe}}}\par

\par The quantum cosmological mini-superspace models analyzed above
included only two degrees of freedom and corresponded to homogeneous
isotropic universes. The inclusion of all degrees of freedom at the
equal footing would present a formidable problem. However, this problem
can be simplified in a perturbative approximation which is a
quantum-mechanical treatment of a perturbed homogeneous isotropic
universe. In particular, the Schr\"odinger equation for gravitons,
with the Hamiltonian equivalent to Eq. (21), can be derived from the
fully quantum cosmological approach as an approximate equation for the
linearized perturbations.\par

\par Let us consider a closed universe governed by an effective
cosmological term~$\Lambda$ and perturbed by weak
gravitational waves. The WD equation for the wave function of this
system can be written in the form (see, for example, Ref. [33], [34]):
\def\XRefId{}
\begin{equation}\SGMPlab\XRefId\vcenter{\halign{\strut\hfil#\hfil&#\hfil\cr
$\displaystyle{{\left\LeftPost{sqb}{{1}\over{2a}}\hskip 0.167em
{{1}\over{a^{p}}}\hskip 0.167em {{%
\SGMPmathgrk{6}}\over{\SGMPmathgrk{6}a}}a^{p}{{\SGMPmathgrk{6}
}\over{\SGMPmathgrk{6}a}}-{{a}\over{2}}+{{%
2}\over{3\SGMPmathgrk{p}}}{\left\LeftPost{par}{{l_{pl}
}\over{l_{0}}}\right\RightPost{par}}^{2}
{{a^{3}}\over{2}}+\sum _{%
nlm}H_{nlm}{\left\LeftPost{par}a,h_{nlm}\right\RightPost{par}}
\right\RightPost{sqb}}\SGMPmathgrk{y}{\left\LeftPost{par}a,{\left\LeftPost{cub}h_{%
nlm}\right\RightPost{cub}}\right\RightPost{par}}=0}$\cr
}}\end{equation}

Here, the h${}_{nlm}$ denotes the amplitude of the gravity wave
perturbation in a given mode (n,l,m). Since we are working in a closed
3{--}space, it is convenient to attribute the indices l,m to
spherical harmonics. The H${}_{nlm}$ denotes the Hamiltonian of the
perturbation:
\(
H_{nlm}=\SGMPmathgrk{p}^{2}_{nlm}/2M+M\SGMPmathgrk{W}^{%
2}_{n}h^{2}_{nlm}/2\),
where $\pi$${}_{nlm}$ is the momentum
canonically conjugated to h${}_{nlm}$, and
\(
M=a^{3},\hskip 0.265em \hskip 0.265em \hskip 0.265em
\hskip 0.265em \hskip 0.265em \SGMPmathgrk{W}_{n}=a^{-1
}{\left\LeftPost{par}n^{2}-1\right\RightPost{par}}^{%
1/2}\).
In what follows we will often omit the indices n,l,m for simplicity.\par

\par The total wave function
\(
\SGMPmathgrk{y}{\left\LeftPost{par}a,\hskip 0.167em
{\left\LeftPost{cub}h\right\RightPost{cub}}
\right\RightPost{par}}\)
 depends on a scale factor
\( a\) and a set of the gravity wave variables
\( h_{nlm}\).
The $\psi$(a, $\{$h$\}$) can be presented in the form
\(
\SGMPmathgrk{y}{\left\LeftPost{par}a,
{\left\LeftPost{cub}h\right\RightPost{cub}}\right\RightPost{par}}
=exp{\left\LeftPost{sqb}-A{\left\LeftPost{par}a\right\RightPost{par}}-A_{%
1}{\left\LeftPost{par}a\right\RightPost{par}}\right\RightPost{sqb}}
\SGMPmathgrk{F}{\left\LeftPost{par}a,
{\left\LeftPost{cub}h\right\RightPost{cub}}\right\RightPost{par}}
\),
where A(a) is the {``}unperturbed{''}
(background) action and A${}_{1}$(a) is the prefactor of the
background wave function.  The $\Phi$(a, $\{$h$\}$) is the part of the total
wave function describing the fluctuations. We assume that  $\psi$(a,
$\{$h$\}$) satisfies the quasiclassical approximation with respect to the
variable a. This allows us to simplify the WD-equation. We assume also
that the fluctuations are weak and do not affect the background so that
the term
\( \SGMPmathgrk{6}^{2}\SGMPmathgrk{F}/\SGMPmathgrk{6}a^{2}\)
can be neglected in Eq.~(61).\par

\par It follows from Eq. (61) that the wave function $\psi$(a, h) for
each mode of fluctuations obeys the Schr\"odinger
equation~
\def\XRefId{}
\begin{equation}\SGMPlab\XRefId\vcenter{\halign{\strut\hfil#\hfil&#\hfil\cr
$\displaystyle{-{{1}\over{i}}\hskip 0.167em {{\SGMPmathgrk{6}
\SGMPmathgrk{y}}\over{a\SGMPmathgrk{6}\SGMPmathgrk{h}}}=H\SGMPmathgrk{y}}$\cr
}}\end{equation}

\noindent
where
\(
\SGMPmathgrk{6}/a\SGMPmathgrk{6}\SGMPmathgrk{h}=-ia^{-1}{\left\LeftPost{par}dA/d
a\right\RightPost{par}}\SGMPmathgrk{6}/\SGMPmathgrk{6}a\),
and
\( H=H_{nlm}\).
The wave function
\(
\SGMPmathgrk{F}{\left\LeftPost{par}a,\hskip 0.167em
{\left\LeftPost{cub}h\right\RightPost{cub}}
\right\RightPost{par}}\)
is constructed as follows:
\(
\SGMPmathgrk{F}=\prod _{nlm}\SGMPmathgrk{y}\).
One can see that in the regime when A(a) describes
classical Lorentzian evolution, that is, when the background space-time
is the De-Sitter solution, Eq. (62) coincides with the Schr\"odinger
equation for the problem considered in Sec. 7. (One has to take into
account some obvious modifications related to the fact that we are
considering now the case k = +1.) However, Eq. (62) has, in fact, a
wider domain of applicability. The assumptions under which Eq. (62) was
derived retain this equation valid in the region where A(a) describes a
classically forbidden behaviour of the universe, i.e. this equation is
valid in the under-barrier region
\( a{\ifmmode<\else$<$\fi}1/H\)
(see Fig. 8) as well. In this region Eq. (62) takes
the form of the Schr\"odinger equation written in the imaginary time.
Thus, the graviton wave function $\psi$(a,h) extends to the classically
forbidden region a{\ifmmode<\else$<$\fi}1/H
and may be sensitive to the form of the
background wave function in this region.\par

\par Our final goal is to show that the initial quantum state of
gravitons at the beginning of the De-Sitter stage (before the parametric
amplification has started) is not unrelated to the form of the
background wave function of the universe in the region
\( a{\ifmmode<\else$<$\fi}1/H\).
Everywhere in our previous discussion we were
assuming that the initial state of gravitons at $\eta$ =
$\eta$${}_{b}$ was the vacuum. The present-day observational
predictions have also been derived under this assumption. However, this
assumption, though quite usual and natural, is not obligatory. If the
initial state of gravitons could have been a non-vacuum state, then it
would lead to the differing predictions for the present day spectrum of
relic gravitons and their squeeze parameters. In this way, by measuring
the actual parameters of relic gravitons, one could learn something
about the wave function of the universe in its classically forbidden
regime.\par

\par One should note, however, that the possible deviations of the
initial quantum state of gravitons from the vacuum state, regardless of
the origin of these deviations, can not be too large. These deviations
should satisfy two requirements. First, they should not violate our
basic assumption that the back-action of gravitons on the background
geometry is always negligibly small. Second, they should not lead to the
predictions for the present day amplitudes which would exceed the
existing experimental limits. By combining these requirements one can
show that only for low-frequency waves and only for cosmological models
with minimally sufficient duration of inflation the initial quantum
state of graviton modes can possibly deviate from the vacuum [35]. In
this case the deviations of the present-day spectrum can be as large as
is shown by the broken line in Fig. 10 for a specific De-Sitter model
with
\(
l_{0}=c/H_{0}=10^{9}\hskip 0.167em l_{%
pl}\).
In this figure the dotted line shows the spectrum
produced from the initial vacuum state in the same model, and the solid
line shows the highest possible inflationary spectrum compatible with
the observational limits. (The solid line is just a low frequency part
of the inflationary spectrum presented in Fig. 4).\par

\par Now we return to the question of which of the background wave
functions are compatable with the deviations of the initial quantum
state of gravitons from the vacuum. As we have already seen, the
Hartle-Hawking wave function $\psi$${}_{HH}$ and the
anti-Hartle-Hawking wave function $\psi$${}_{aHH}$ are, in a sense,
two extremes in the description of the classically forbidden domain.
Each of these extremes can be used in Eq. (62) as a background wave
function. For each of them the solution to Eq. (62) can be presented in
the Gaussian form (compare with Eq. (15))
\(
\SGMPmathgrk{y}{\left\LeftPost{par}h,
\SGMPmathgrk{h}\right\RightPost{par}}=C{\left\LeftPost{par}
\SGMPmathgrk{h}\right\RightPost{par}}
e^{-B{\left\LeftPost{par}\SGMPmathgrk{h}\right\RightPost{par}}
h^{2}}\).
Here
\(
B{\left\LeftPost{par}\SGMPmathgrk{h}\right\RightPost{par}}\)
is constricted as a linear combination of two
independent complex solutions to the classical wave equation (6), and
the vacuum state at
\(
\SGMPmathgrk{h}=\SGMPmathgrk{h}_{b}\)
corresponds to the value
\(
B{\left\LeftPost{par}\SGMPmathgrk{h}\right\RightPost{par}}=
B{\left\LeftPost{par}\SGMPmathgrk{h}_{%
b}\right\RightPost{par}}\),
where
\(
B{\left\LeftPost{par}\SGMPmathgrk{h}_{b}\right\RightPost{par}}={{%
1}\over{2}}\SGMPmathgrk{w}_{b},\hskip 0.167em \SGMPmathgrk{w}_{%
b}={\left\LeftPost{par}n^{2}-1\right\RightPost{par}}^{%
{{1}\over{2}}}a^{-2}{\left\LeftPost{par}
\SGMPmathgrk{h}_{b}\right\RightPost{par}}\).
It is important that the choice of
$\psi$${}_{HH}$ or $\psi$${}_{aHH}$ in the classically forbidden
region restricts the function
\(
B{\left\LeftPost{par}\SGMPmathgrk{h}\right\RightPost{par}}\)
in different ways if one is willing to subject the wave function
\(
\SGMPmathgrk{y}{\left\LeftPost{par}h,
\hskip 0.167em \SGMPmathgrk{h}\right\RightPost{par}}
\)
to the condition of normalizability:
\(
\int _{-\infty }^{\infty }\SGMPmathgrk{y}^{%
*}\SGMPmathgrk{y}dh{\ifmmode<\else$<$\fi}\infty \)
[34]. If this condition is imposed, it requires Re
\(
B{\left\LeftPost{par}\SGMPmathgrk{h}\right\RightPost{par}}
{\ifmmode>\else$>$\fi}0\).
It turns out that this requirement singles out the
vacuum initial state if the background wave function is
$\psi$${}_{HH}$ and it leaves room for non-vacuum initial states if
the background wave function is $\psi$${}_{aHH}$. Thus, if relic
gravitational waves are detected with properties different from those
following from the initial vacuum state one could conclude that the
universe was described by $\psi$${}_{aHH}$, and not by
$\psi$${}_{HH}$, in the classically forbidden regime. This would
strengthen the hypothesis that the universe was created in a quantum
process similar to quantum tunneling or decay. Thus, the difference
between possible wave functions of the universe in the classically
forbidden regime can be distinguished by exploring the properties of the
gravitational wave background existing now.\SGMPnewline
\par

\par \underline{{\protect\normalsize {\bf Acknowlegements}}}\par

\par I acknowledge numerous discussions and collaborations with
G.~W.~Gibbons, L.~V.~Rozhansky, Yu.~V.~Sidorov and late Ya.~B.~Zeldovich
which became the basis for these lectures. This paper is a significantly
revised and updated version of the manuscript which was originally
prepared and distributed in a xerox form as a set of lectures for VI-th
Brasilian School on Cosmology and Gravitation (Rio de Janeiro, July,
1989). The work on both versions of the paper has been done during my
two visits to the University of British Columbia and the Canadian
Institute for Advanced Research. I am grateful to Bill Unruh for many
discussions and warm hospitality. My thanks also go to J.~Maxwell,
A.~Artuso, B.~Wiseman and D.~Bruce for typing the manuscript.\par

\newpage

\par \underline{{\protect\normalsize {\bf References}}}
\SGMPbeginList{numeral}{}
\SGMPitem\def\XRefId{}\SGMPlab\XRefId Ya. B. Zeldovich and I.D. Novikov.
 Structure and Evolution of the
Universe. Nauka, Moscow, 1975; University of Chicago Press, Chicago,
1983.\SGMPnewline
S. Weinberg. Gravitation and Cosmology. Wiley, N.Y.,
1972.\SGMPnewline
P.J.E. Peebles. Physical Cosmology. Princeton University Press,
1971.\SGMPnewline
J.E. Gunn, M.S. Longair, and M.J. Rees. Observational Cosmology.
SAAS-FEE, 1978.
\SGMPitem\def\XRefId{}\SGMPlab\XRefId G.F. Smoot et al.
(to be published in Astrophys. J. Lett).
\SGMPitem\def\XRefId{}\SGMPlab\XRefId B.R. Harrison, Phys. Rev.
\underline{D1}, 2726
(1970).\SGMPnewline
Ya. B. Zeldovich, Mon. Not. R. Astron. Soc.
\underline{180}, 1 (1972).
\SGMPitem\def\XRefId{}\SGMPlab\XRefId A. Guth, Phys. Rev. \underline{D23},
389 (1981).\SGMPnewline
A.D. Linde, Rep. Progr. Phys.
\underline{47}, 925 (1984); Particle
Physics and Inflationary Cosmology. Chur, Switzerland, Harwood,
1990.\SGMPnewline
M.S. Turner, In: Proceed, Cargese School on Fundam. Physics and
Cosmology, eds. J. Audouze, J. Tran Thanh Van; Ed. Front:
Gif-Sur-Yvette, 1985.
\SGMPitem\def\XRefId{}\SGMPlab\XRefId S.W. Hawking, G. Ellis.
The Large Scale Structure of Space-Time.
Cambridge University Press, 1973.
\SGMPitem\def\XRefId{}\SGMPlab\XRefId L.P. Grishchuk, Ya. B.,
Zeldovich. In: Quantum Structure of Space
and Time. eds. M. Duff, C. Isham. Cambridge University Press, 1982, p.
409.\SGMPnewline
L.P. Grishchuk, Ya. B. Zeldovich. In: Ya. B. Zeldovich, Selected Works:
Particles, Nuclei and the Universe. (in Russian), Nauka, Moscow,
1985.
\SGMPitem\def\XRefId{}\SGMPlab\XRefId S.W. Hawking. Phys. Rev.
\underline{D37}, 904
(1988).\SGMPnewline
S. Coleman. Nucl. Phys. \underline{B307},
867 (1988).\SGMPnewline
G.V. Lavrelashvili, V.A. Rubakov, and P.G. Tinyakov. JETP Lett.
\underline{46}, 167 (1987); Nucl. Phys.
\underline{B299}, 757
(1988).\SGMPnewline
S. Giddings and A. Strominger. Nucl. Phys.
\underline{B307}, 854
(1988).\SGMPnewline
W.G. Unruh. Phys. Rev. \underline{D40},
1053 (1989).
\SGMPitem\def\XRefId{}\SGMPlab\XRefId L.P. Grishchuk. Sov. Phys. JETP,
\underline{40}, 409 (1974); Lett. Nuovo
Cimento \underline{12}, 60 (1975); JETP
Lett. \underline{23}, 293 (1976); Ann. N.Y.
Acad. Sci. \underline{302}, 439 (1977);
Sov. Phys. Uspekhi \underline{20}, 319
(1977).
\SGMPitem\def\XRefId{}\SGMPlab\XRefId L.P. Grishchuk, Sov. Phys. Uspekhi
\underline{31}, 940 (1989); In: Proceed. of
11th Intern. Conf. GRG, ed. M. MacCallum. Cambr. Univ. Press
1987.
\SGMPitem\def\XRefId{}\SGMPlab\XRefId A.A. Starobinsky. JETP. Lett.
\underline{30}, 682 (1979).
\SGMPitem\def\XRefId{}\SGMPlab\XRefId V.A. Rubakov, M.V. Sazhin,
and A.V. Veryashin. Phys. Lett.
\underline{B115}, 189 (1982).
\SGMPitem\def\XRefId{}\SGMPlab\XRefId R. Fabbri and M.D. Pollock. Phys. Lett.
\underline{B125}, 445
(1983).\SGMPnewline
L.F. Abbott and M.B. Wise. Nucl. Phys.
\underline{B224}, 541
(1984).\SGMPnewline
E. Witten. Phys. Rev. \underline{D30}, 272
(1984).\SGMPnewline
L.F. Abbott and D.D. Harari. Nucl. Phys.
\underline{B264}, 487
(1986).\SGMPnewline
L.M. Krauss. Gen. Rel. and Gravit.
\underline{18}, 723 (1986).
\SGMPitem\def\XRefId{}\SGMPlab\XRefId L.P. Grishchuk and M. Solokhin. Phys.
Rev.
\underline{D43}, 2566 (1991).
\SGMPitem\def\XRefId{}\SGMPlab\XRefId B.L. Schumaker. Phys. Repts.
\underline{135}, 317 (1986).
\SGMPitem\def\XRefId{}\SGMPlab\XRefId L.P. Grishchuk,
In: Workshop on Squeezed States and Uncertainty
Relations. Eds. D. Han, Y.S. Kim and W.W. Zachary, NASA Conference
Publication 3135, 1992, p. 329.
\SGMPitem\def\XRefId{}\SGMPlab\XRefId L.P. Grishchuk, H.A.
Haus and K. Bergman. Phys. Rev.
\underline{D46}, N4, 15 August,
1992.
\SGMPitem\def\XRefId{}\SGMPlab\XRefId L.P. Grishchuk,
Yu. V. Sidorov. Class. Quant. Grav. (Letters)
\underline{6}, L161, 1989; Phys. Rev.
\underline{D42}, 3413 (1990).
\SGMPitem\def\XRefId{}\SGMPlab\XRefId L.P. Grishchuk. In:
 Proceedings of the MG6 conference. Kyoto,
Japan, 1991. ed. T. Nakamura. World Scientific, 1992.
\SGMPitem\def\XRefId{}\SGMPlab\XRefId V.F. Mukhanov,
H.A. Feldman and R.H. Brandenberger. Phys. Reports,
\underline{215}, N5, 6, 1992.
\SGMPitem\def\XRefId{}\SGMPlab\XRefId B. DeWitt. Phys. Rev.
\underline{160}, 1113
(1967).\SGMPnewline
J.A. Wheeler, In: Batelle Recontres. eds. C. DeWitt and J.A. Wheeler,
Benjamin, 1988.\SGMPnewline
S.W. Hawking. In: 300 Years of Gravitation. eds. S.W. Hawking and W.
Israel. Cambr. Univ. Press 1988.\SGMPnewline
J.B. Hartle In: Proceed. of the Cargese 1986 Summer Inst.; eds. J.
Hartle and B. Carter, Plenum, 1987.
\SGMPitem\def\XRefId{}\SGMPlab\XRefId S.W. Hawking, Nucl. Phys.
\underline{B239}, 257
(1984).\SGMPnewline
S.W. Hawking and D. Page. Nucl. Phys.
\underline{B264}, 185 (1986).
\SGMPitem\def\XRefId{}\SGMPlab\XRefId A. Vilenkin. Phys. Rev.
\underline{D37}, 888
(1988).\SGMPnewline
J.J. Halliwell. Quantum Cosmology: an Introductory Review. Preprint
NSF-ITP-88{--}131, 1988.
\SGMPitem\def\XRefId{}\SGMPlab\XRefId V.A. Belinsky, L.P. Grishchuk,
I.M. Khalatnikov, and Ya. B.
Zeldovich. Phys. Lett. \underline{B155},
232 (1985); Sov. Phys. JETP \underline{62},
195 (1985).
\SGMPitem\def\XRefId{}\SGMPlab\XRefId L.P. Grishchuk and L.V. Rozhansky. Phys.
Lett.
\underline{B234}, 9 (1990).
\SGMPitem\def\XRefId{}\SGMPlab\XRefId L.P. Grishchuk and
Yu. V. Sidorov. Sov. Phys. JETP
\underline{67}, 1533 (1989); L.P.
Grishchuk. Quantum cosmology: from the space of classical solutions to
the space of wavefunctions. In: Proceed. Friedman Conference. W.S.
Singapore, 1990, p. 289.
\SGMPitem\def\XRefId{}\SGMPlab\XRefId A. Vilenkin. Nucl. Phys.
\underline{B252}, 141 (1985).
\SGMPitem\def\XRefId{}\SGMPlab\XRefId
 G.W. Gibbons and L.P. Grishchuk. Nucl. Phys.
\underline{B313}, 736 (1989).
\SGMPitem\def\XRefId{}\SGMPlab\XRefId G.W. Gibbons,
 S.W. Hawking and J.M. Stewart. Nucl Phys.
\underline{B281}, 736 (1987).
\SGMPitem\def\XRefId{}\SGMPlab\XRefId J.B. Hartle and
S.W. Hawking. Phys. Rev.
\underline{D28}, 2960 (1983).
\SGMPitem\def\XRefId{}\SGMPlab\XRefId L.P. Grishchuk and
L.V. Rozhansky. Phys. Lett.
\underline{B208}, 369 (1988).
\SGMPitem\def\XRefId{}\SGMPlab\XRefId M.S. Turner. Phys. Rev.
\underline{D44}, 3737 (1991).
\SGMPitem\def\XRefId{}\SGMPlab\XRefId L.P. Grishchuk. Phys. Rev.
\underline{D45}, 4717 (1992).
\SGMPitem\def\XRefId{}\SGMPlab\XRefId J.J. Halliwell and S.W. Hawking. Phys.
Rev.
\underline{31}, 1777 (1985).
\SGMPitem\def\XRefId{}\SGMPlab\XRefId S. Wada. Nucl. Phys.
\underline{B276}, 729 (1986).
\SGMPitem\def\XRefId{}\SGMPlab\XRefId L.P. Grishchuk and
Yu. V. Sidorov. Class. Quantum Grav.
\underline{6}, L155 (1989).
\SGMPendList
\par

\newpage

\par \underline{{\protect\normalsize {\bf Figure
Captions}}}
\SGMPbeginList{numeral}{}
\SGMPitem\def\XRefId{}\SGMPlab\XRefId The scale factor
 of a complete cosmological theory.
\SGMPitem\def\XRefId{}\SGMPlab\XRefId Parametric (superadiabatic)
 amplification of waves.
\SGMPitem\def\XRefId{}\SGMPlab\XRefId The potential
\(
U{\left\LeftPost{par}\SGMPmathgrk{h}\right\RightPost{par}}\)
 for the inflationary {---} radiation-dominated
{---} matter-dominated cosmological model.
\SGMPitem\def\XRefId{}\SGMPlab\XRefId Theoretical predictions
and experimental limits for stochastic
gravitational waves.
\SGMPitem\def\XRefId{}\SGMPlab\XRefId Variancies for coherent
and squeezed quantum states.
\SGMPitem\def\XRefId{}\SGMPlab\XRefId Classical trajectories
at the compactified
\(
{\left\LeftPost{par}\SGMPmathgrk{f},\hskip 0.167em
\MthAcnt {\SGMPmathgrk{f}}{\dot }
\right\RightPost{par}}\)
 phase diagram.
\SGMPitem\def\XRefId{}\SGMPlab\XRefId Classical paths in the
\(
{\left\LeftPost{par}\SGMPmathgrk{a},\hskip 0.167em
\SGMPmathgrk{f}\right\RightPost{par}}
\)
 configuration space.
\SGMPitem\def\XRefId{}\SGMPlab\XRefId The potential
\(
V{\left\LeftPost{par}a\right\RightPost{par}}\)
{}.
\SGMPitem\def\XRefId{}\SGMPlab\XRefId Two branches of
the caustic line and its singularity.
\SGMPitem\def\XRefId{}\SGMPlab\XRefId Possible spectra
of relic gravitons.
\SGMPendList
\par

\end{document}